\documentclass[final,5p,times,twocolumn]{elsarticle}
\IfFileExists{todonotes.sty}{%
  \usepackage{todonotes}%
}{%
  \newcommand{\todo}[2][]{}
  \newcommand{\missingfigure}[2][]{}
}

\usepackage{amsmath,amssymb,amsfonts,amsthm} % Provides mathematical symbols, fonts, and theorem environments

\usepackage{algorithm,algorithmicx,algpseudocode} % Supports writing algorithm pseudocode

\usepackage{graphicx,epstopdf,float,tikz,pgfplots} % Handles images, EPS support, drawings, and plots
\usepackage{subcaption}

\usepackage{makecell,multirow,tabularx,booktabs,arydshln} % Enhances table functionality, supports merged cells and dashed lines

\usepackage{cite} % Manages citations efficiently and optimizes reference display

\usepackage{xcolor} % Enables colored text
\usepackage{latexsym,enumerate,soul} % Special symbols, custom enumerations, and underlining

\usepackage{mathrsfs,exscale,relsize} % Additional math fonts and scalable symbols

\usepackage[bookmarks=false]{hyperref} % Enables hyperlinks but disables bookmarks

\makeatletter
\g@addto@macro\UrlBreaks{\do\-\do\_}
\makeatother
\definecolor{lavendergray}{rgb}{0.77, 0.76, 0.82}

\def\BibTeX{{\rm B\kern-.05em{\sc i\kern-.025em b}\kern-.08em
T\kern-.1667em\lower.7ex\hbox{E}\kern-.125emX}}

\newtheorem{proposition}{Proposition}
\newtheorem{corollary}{Corollary}

\newcommand{\revise}[1]{{\color{black}{#1}}}

\newcommand{\revisenew}[1]{{\color{black}{#1}}}

\pgfplotsset{compat=1.18}

\journal{Computer Networks}

\begin{document}
\begin{frontmatter}
%% Title
\title{\revise{Energy-Aware} Holistic Optimization in UAV-Assisted Fog Computing: Attitude, Trajectory, and Task Assignment}

%% Authors
\author[uic,bu]{Shuaijun Liu\fnref{website}}
\ead{shuaijun@bu.edu}
\author[uw]{Jinqiu Du}
\ead{turbodddu@gmail.com}
\author[columbia]{Yaxin Zheng}
\ead{kristenzyx@163.com}
\author[sysu]{Jiaying Yin}
\ead{yinjy35@mail.sysu.edu.cn}
\author[uic]{Yuhui Deng}
\ead{ivandeng@bnbu.edu.cn}
\author[uic]{Jingjin Wu\texorpdfstring{\corref{cor1}}{}}
\ead{jj.wu@ieee.org}

%% Corresponding author
\cortext[cor1]{Corresponding author: Jingjin Wu.}

%% Affiliations
\affiliation[uic]{organization={Guangdong Provincial Key Laboratory of IRADS, Beijing Normal-Hong Kong Baptist University}, 
            city={Guangdong}, 
            country={China}}
\affiliation[bu]{organization={Department of Computer Science, Boston University}, 
            city={Boston}, 
            state={MA}, 
            country={USA}}
\affiliation[uw]{organization={Department of Biostatistics, University of Washington}, 
            city={Seattle}, 
            state={WA}, 
            country={USA}}
\affiliation[columbia]{organization={Department of Statistics, Columbia University}, 
            city={New York}, 
            state={NY}, 
            country={USA}}
\affiliation[sysu]{organization={Institute of Precision Medicine, The First Affiliated Hospital, Sun Yat-Sen University}, 
            city={Guangdong}, 
            country={China}}

%\affiliation[gdlab]{organization={Guangdong Provincial Key Laboratory of Interdisciplinary Research and Application for Data Science}, 
  %          city={Guangdong}, 
   %         country={China}}
            
%% Acknowledgments
\tnotetext[funding]{This work was partly supported 
the Guangdong Provincial Key Laboratory of Interdisciplinary Research and Application for Data Science, BNU-HKBU United International College (Project code 2022B1212010006), 
and Guangdong Higher Education Upgrading Plan (2021-2025) UIC R0400001-22 and R0400024-22.}

\tnotetext[conference]{A preliminary version of this paper was presented at IEEE HPCC 2022 in December 2022, and was included in its proceedings (DOI: \href{http://dx.doi.org/10.1109/hpcc-dss-smartcity-dependsys57074.2022.00217}{10.1109/hpcc-dss-smartcity-dependsys57074.2022.00217}).}
\fntext[website]{Project website: \textcolor{blue}{\url{https://shuaijun-liu.github.io/HATTO-UFog}}.}

%% Abstract
\begin{abstract}
Unmanned Aerial Vehicles (UAVs) have significantly enhanced fog computing by acting as both flexible computation platforms and communication mobile relays. 
In this paper, we consider \revisenew{four} important and interdependent modules: attitude control, trajectory planning, \revise{resource allocation}, and task assignment, and propose a holistic framework that 
jointly optimizes the total latency and energy consumption for UAV-assisted fog computing in a three-dimensional spatial domain with varying terrain elevations and dynamic task generations. 
We first establish a fuzzy-enhanced adaptive reinforcement proportional-integral-derivative control model to control the attitude. 
Then, we propose an enhanced Ant Colony System (ACS) based algorithm, that includes a safety value and a decoupling mechanism to overcome the convergence issue in classical ACS, to compute the optimal UAV trajectory. 
Finally, we design an algorithm based on the Particle Swarm Optimization technique, to determine where each offloaded task should be executed. 
Under our proposed framework, the outcome of one module would affect the decision-making in another, providing a holistic perspective of the system and thus leading to improved solutions. 
We demonstrate by extensive simulation results that our proposed framework can significantly improve the overall performance, measured by latency and energy consumption, compared to existing mainstream approaches.
\end{abstract}

%% Keywords
\begin{keyword}
Fog Computing \sep Unmanned Aerial Vehicles (UAV) \sep Attitude Control \sep Trajectory Planning \sep Ant Colony Algorithm
\end{keyword}

%%Research highlights
% \begin{highlights}
% \item Research highlight 1
% \item Research highlight 2
% \end{highlights}

\end{frontmatter}

\vspace{-0.15cm}  % *****
\section{Introduction}
Driven by the development of the Internet of Things (IoT)~\citep{Ashton1999}, mobile terminal devices such as smartphones and tablets are now capable of generating and collecting massive amounts of data. However, the ability of processing these data, such as performing computational tasks, in the IoT devices are still limited \citep{Hu2021,pei2022uav}. 
On the other hand, Unmanned Aerial Vehicles (UAVs) have been recently identified as a versatile platform that connects IoT devices and servers or data centers via the network edge~\citep{Ghdiri2021}. In addition, some UAVs are equipped with computational capabilities and thus can be regarded as ``moving fog nodes" for offloading certain computational tasks. To fully utilize the versatility and flexibility of UAVs in fog computing, key considerations include the management of each UAV's attitude, the strategic planning of their respective trajectories, and the efficient task assignment policy to determine the appropriate computing device for each task.

We consider a fog computing environment with a single \revise{UAV} deployed at the network edge, aiming at maximizing the energy efficiency by collaboratively controlling the attitudes, planning the trajectories, allocating the transmission and computation resources, and assigning computing tasks to appropriate devices for execution. Specifically, \textbf{attitude control} (by adjusting pitch, roll, and yaw) ensures a stable and precise orientation of the UAV during operation, which is a necessary condition to maintain a high quality of communication with IoT devices and other fog nodes \citep{Poksawat2018}.  
In addition, a stable attitude facilitates the UAV to effectively perform computation and storage tasks as a fog node. On the other hand, \textbf{trajectory planning} of the UAV  can reduce its power consumption by identifying the most efficient path to collect data and tasks based on the locations of IoT devices~\citep{Besada-Portas2010}. Finally, \textbf{task assignment} refers to the process of deciding whether a specific task should be handled locally by the IoT device, processed in the fog layer (including the UAV), or offloaded to the central cloud based on real-time application-specific scenarios~\citep{rottondi2021scheduling}. Specific objectives, including minimizing latency, maximizing throughput, or optimizing energy efficiency, can be achieved by assigning tasks to appropriate devices. Task assignment is often jointly optimized with \textbf{resource allocation}, where transmission resources such as power and bandwidth are distributed among different transmission pairs in the network, to facilitate the transmission process and improve the overall efficiency. 

The four processes that we consider are inherently linked in UAV-assisted fog computing. For example, a UAV's attitude control would ensure that it maintains optimal orientations while following a planned trajectory or processing a task. Also, when deciding whether to offload a certain task to the fog or the cloud and how to allocate relevant transmission resources, the energy consumption and latency for a certain UAV to reach the proximity of the IoT initiating the task along a planned trajectory should also be taken into account.

Existing studies have considered two or three processes for joint optimization. For example, Cheng \textit{et al.} \citep{cheng2022deep} proposed three decision-making algorithms to solve the joint optimization problem involving energy consumption and mean delay. \revise{Zhou \textit{et al.}~\citep{Zhou2022two} proposed a two–time-scale optimization framework that jointly determines caching placement and task offloading decisions while adaptively adjusting the UAV trajectory.} However, few  considered all these aspects together in an interconnected manner. A summary of relevant studies is provided in Table~\ref{tab:related}, and we will discuss them in more detail in Section~\ref{sec:rw}.

\revise{This work substantially extends our earlier conference paper~\citep{Liu2022}, to better capture the practical dynamics of UAV-assisted fog computing networks. The major improvements include: 1) While the previous study focused on a two-dimensional setting with simplified flight and communication assumptions, the present work develops a three-dimensional, terrain-aware network model that incorporates altitude variation and realistic environmental constraints; 2) We introduce an additional module (attitude control)  for quadrotor UAVs, enabling the system to maintain stable flight, enhance link reliability under varying orientations, and mitigate latency during altitude transitions; and 3) We refine the algorithm for trajectory planning to better avoid deadlock by introducing decoupling and safety values mechanisms. These extensions allow the UAV to adapt its physical configuration and communication behavior in a coordinated manner, thereby improving overall energy efficiency, communication stability, and task execution performance beyond what was achieved in~\citep{Liu2022}.} Compared to similar existing studies (e.g.,\citep{Tan2024TD3,Zheng2024GASCA}), which only considered the trajectory planning and flight attitude at independent static points, we jointly consider planning the optimal trajectory and determining the attitudes, taking into account the extra consumption required for changing attitude along the trajectory. Our research is expected to provide new useful insights in UAV-assisted fog computing applications for improving the overall performance and efficiency. 

\revise{Although some of the individual strategies employed in this work, such as fuzzy, adaptive, and PID algorithms, are well-established, the novelty of our approach lies in how these algorithms are integrated and jointly optimized within a unified UAV-assisted fog computing framework. In contrast to most existing studies that treat control and networking processes separately, our framework co-designs attitude control, trajectory planning, task assignment, and resource allocation in a mutually dependent manner. This cross-layer integration enables the UAV to adapt its physical dynamics and communication decisions in real time according to network conditions, task demands, and environmental factors. As a result, the proposed approach improves system-level performance in terms of energy efficiency, communication reliability, and latency, thereby providing a new perspective on the joint optimization of control and communication functions in UAV-enabled fog computing networks.}

%\vspace{0.15cm} % *****

The contributions of this paper are summarized as follows.
\begin{itemize}
    \item \revise{From a computer and communication network perspective, this work proposes a unified optimization framework for UAV-assisted fog computing networks that jointly integrates attitude control, trajectory planning, task assignment, and resource allocation within a three-dimensional network topology. By coupling the UAV’s physical-layer control (attitude control) with network-layer decision making (resource allocation and task assignment), our cross-layer joint optimization framework captures the mutual dependencies among communication stability, mobility dynamics, and computational load distribution, thereby achieving adaptive connectivity, reduced latency, and improved overall network performance in realistic, terrain-aware IoT environments.}
    
    \item We develop a fuzzy-enhanced adaptive reinforcement proportional-integral-derivative (FEAR-PID) model for attitude management of quadrator UAVs. Compared with classical PID and conventional fuzzy-PID control commonly used in existing studies, FEAR-PID captures interdependencies among parameters and adapts dynamically to environmental changes. Our results demonstrate that FEAR-PID significantly enhances stability during takeoff, cruising, and landing phases, thereby effectively reducing latency and energy consumption in UAV-assisted fog computing systems.

    \item We propose a computationally efficient algorithm called ACS-DS (Ant Colony System  with Decoupling and Safety values) for UAV trajectory planning. The ACS-DS algorithm integrates decoupling and safety value mechanisms to address typical limitations of classical ACS, such as slow convergence rates and susceptibility to local optima. Numerical experiments also confirm ACS-DS's superior convergence performance relative to mainstream heuristic and reinforcement learning-based methods.

    \item We propose a heuristic algorithm based on PSO principles to effectively resolve the resource allocation and task assignment problems, given that an initial trajectory of the UAV has been determined. The heuristic algorithm efficiently overcomes the inherent complexities of the underlying non-convex optimization problems, providing quasi-optimal solutions for task assignment and resource allocation decisions.

    \item We demonstrate, through extensive numerical experiments, that our proposed holistic framework can reduce overall operational efficiency cost by more than 67\% reduction in overall operational costs compared to existing heuristic and reinforcement learning-based methodologies. Our analysis and results underscore the importance of considering interdependencies among attitude control, trajectory planning, resource allocation, and task assignment in UAV-assisted fog computing. Consequently, the holistic approach significantly outperforms methods optimizing individual components separately, highlighting the cumulative benefits of joint optimization in such environment.
\end{itemize}

The rest of this paper is organized as follows. Section~\ref{sec:rw} reviews recent advancements on attitude control, trajectory planning, and task assignment in UAV-assisted fog computing. Section~\ref{subsec: Network} provides descriptions on the UAV architecture as well as key metrics at the system level. Section~\ref{sec:Joint optimization model} explains the formulation of the joint optimization problem. Section~\ref{sec: Algorithm for Trajectory} describes the proposed computationally efficient algorithms to solve the problem in detail. Section~\ref{sec: experiment} demonstrates the improvements of the proposed algorithm by extensive numerical results. Section~\ref{sec: conclusion}~concludes the paper.

%%%%%%%%%%%%%%%%%%%%%%%%%%%%%%%%%%%%%%%%%%%

\section{Related Work}
\label{sec:rw}

%The fuzzy proportional-integral-derivative (PID) control system was adopted in attitude control and trajectory planning processes~\citep{Obias2019,Batikan2016}. It has been demonstrated that a proper combination of proportional-integral (PI) and proportional-derivative (PD) values and an optimally tuned fuzzy gain can better stabilize and control the attitude of quadrotor UAVs, as well as track their trajectories with smaller errors~\citep{Yu2024tvt}. 
\subsection{Attitude Control}
The proportional-integral-derivative (PID) control system is a fundamental approach widely used in the attitude control and trajectory planning of quadrotor UAVs~\citep{Obias2019,Batikan2016}. Building on this foundation, fuzzy PID control introduces adaptive capabilities by incorporating fuzzy logic to dynamically adjust the proportional-integral (PI) and proportional-derivative (PD) components, resulting in improved stability and reduced trajectory tracking errors ~\citep{Yu2024tvt}. Recent studies further enhance these methods by leveraging reinforcement learning (RL). For instance, RL has been employed to regulate linear velocity while fuzzy logic manages angular velocity, achieving a complementary control strategy~\citep{Xia2024}. Another approach is the participation of RL in the construction of the PID controllers, which optimizes the gain parameters in real-time to achieve more accurate adaptive control under dynamic and complex environmental conditions~\citep{Shri2024}.

\subsection{Trajectory Planning}

For trajectory planning of UAVs, heuristic algorithms such as PSO \citep{Shen2022}, ACS \citep{diallo2019efficient,Yan2024CPSACO}, and genetic algorithm (GA) \citep{Zheng2024GASCA}, have been adopted to overcome the space and computation complexities in such problems. Compared with another branch of approaches that use neural networks and deep learning as the key techniques (e.g.,~\citep{Tan2024TD3,tao2021hybrid}), heuristic algorithms are more interpretable and less data dependent, and thus more appropriate for UAV-assisted fog computing scenarios where the operational environments are usually highly diverse and dynamic \citep{Liu2024Meterological}, and transparent heuristics are preferred for regulators to validate and verify the operation. Among the heuristic algorithms, the PSO \citep{jain2022overview} is one of the most commonly used techniques in such problems. However, one major concern of applying PSO in complex systems is that PSO may be converged prematurely to local optima~\citep{shin2020uav}.

ACS, based on the study of ants searching for food, is another commonly adopted technique in discrete and continuous optimization problems~\citep{dorigo1999ant,socha2004aco}. While ACS-based approaches are well-known for their robustness, they also suffer the disadvantage of being prone to local optima due to premature convergence like PSO. To overcome this issue, Wang \textit{et al.}~\citep{Wang2021} presented an adaptive double-layer ant colony optimization algorithm (DL-ACS) based on an elitist strategy (ADAS) and an improved moving average algorithm (IMA) to solve a three-dimensional UAV trajectory planning problem. Recently, Yan \textit{et al.}~\citep{Yan2024CPSACO} proposed a chaotic-polarized-simulated ant colony optimization (CPS-ACO) algorithm, by incorporating chaotic mapping for initial pheromone distribution, a polarizing pheromone recording rule, and a simulated annealing mechanism, CPS-ACO demonstrated enhanced convergence speed and robustness against local optima in trajectory planning.

\subsection{Task Assignment}
Resource allocation or task assignment problems have been extensively studied in existing studies. For example, Cui \textit{et al.}~\citep{Cui2020} proposed a multi-agent Q-learning-based reinforcement learning (MARL) framework, where each agent independently executes the allocation algorithm to optimize the overall energy efficiency. Li \textit{et al.}~\citep{li2020energy} both considered a joint optimization problem involving trajectory optimization and task allocation, with the goal of minimizing UAV energy consumption and optimizing computation offloading and using successive convex approximation (SCA) technique to solve it. Wu \textit{et al.}~\citep{Wu2023} proposed a cooperative multi-agent deep reinforcement learning framework, which combines task assignment and allocation of limited communication resources to minimize the overall energy consumption and delay.

\subsection{Joint Optimization}

One notable issue of applying ACS, GA or PSO in joint optimization involving trajectory planning and task assignment is that the algorithm may enter the deadlock state where one or more tasks wait endlessly for resources~\citep{Francesca2011}. Hou \textit{et al.}~\citep{HOU2022} proposed ACS-based algorithms with enhanced communication mechanisms to avoid deadlocks. While the proposed methods managed to reduce the likelihood of deadlocks, they did not eliminate the possibilities of such undesirable events. Another effort to overcome the deadlock problem is segmented planning and reintegration~\citep{Liu2022IOT}, which introduces self-healing mechanisms like In-Road Repairing and Intersection Repairing. These methods dynamically recalculate paths and penalize deadlocked routes using a negative reinforcement strategy. While this approach effectively avoids deadlocks and improves reliability, it incurs computational overhead due to frequent evaluations and path adjustments. %Zheng \textit{et al.}~\citep{Zheng2023aco} proposed to solve the problem by optimization heuristic design, introducing a backoff mechanism and a pheromone penalty mechanism to effectively alleviate the deadlock and local optima problem by preventing the selection of suboptimal paths. This approach coincides with the ideas presented in this study and provides valuable insights for our research.

Another emerging approach to solving UAV-related planning and optimization problems is deep reinforcement learning (DRL) (e.g.,~\citep{Qin2023Multi,Wei2023Joint,Zheng2024TD3}). Recently, Tan \textit{et al.}~\citep{Tan2024TD3} built a two-layer optimization framework by combining differential evolutionary algorithms and deep reinforcement learning to optimize the UAV deployment, tasks assignment and resource allocation efficiency when participating in multi-access edge computing. \revise{Proximal policy optimization-based deep reinforcement learning algorithms were used in~\citep{Huang2023online} to learn the optimal offloading strategy for dependent tasks in a joint optimization problem that also involves trajectory planning. DRL-based approaches have been also jointly applied with other optimization modules in edge computing, such as blockchain consensus leader election and waiting time window decision~\citep{Xu2025dynamic}, and estimation of resource requirements through digital twins~\citep{Xu2023digital}.}

Compared with DRL-based approaches, the heuristic algorithms (such as ACS and PSO) are known to be more scalable, more adaptive and less computationally intensive~\citep{Yew2023investigating}. These advantages make them particularly well-suited for the dynamic and large-scale nature of UAV-assisted fog computing scenarios. We will also compare their performance and computational efficiency in Section~\ref{sec: experiment} of this paper.

\begin{table*}[ht]
\caption{A SUMMARY AND COMPARISON OF RELATED STUDIES}
\centering
\scalebox{0.8}{
\begin{tabular}{lllll}
\toprule 
\textbf{Research} & \textbf{Year} & \textbf{Process(es) focused} & \textbf{Method(s)} & \textbf{Objective(s)} \\
\midrule
\textbf{\citep{Dev2022}} & 2022 & TA & Heuristic Harris Hawks Optimization  & Min. energy consumption \\
\textbf{\citep{Shen2022}} & 2022 & TP & Clustered-PSO & Min. energy efficiency\\
\textbf{\citep{HOU2022}} & 2022 & TP & Enhanced ACS & Avoid path deadlock \\
\textbf{\citep{Liu2022IOT}} & 2022 & TP & Segmented ACS with Self-healing Routing & Avoid path deadlock\\
\textbf{\citep{Wang2021}} & 2021 & 3D-TP & Double-layer ACS & Optimal trajectory \\
\textbf{\citep{Zheng2024GASCA}} & 2024 & 3D-TP & Genetic Algorithm (GA) \& SCA & Optimal trajectory \\
\textbf{\citep{Yan2024CPSACO}} & 2024 & 3D-TP & Chaotic-Polarized-Simulated scheme ACO & Min. trajectory loss rate \\
\textbf{\citep{Obias2019}} & 2019 & ATC & Traditional PID controller & Max. control accuracy \\
\textbf{\citep{Yu2024tvt}} & 2024 & ATC & Fuzzy PID Controller & Min. control bias \\
\textbf{\citep{Shri2024}} & 2024 & ATC & RL (Deep Q-Network) PID Controller & Max. control accuracy \\
\textbf{\citep{Wu2023}} & 2023 & TA \& RA & MARL \& Stochastic Game Model & Min. energy consumption and latency \\
\textbf{\citep{Batikan2016}} & 2016 & 3D-TP \& ATC & Fuzzy PID controller & Control attitude and track trajectory \\
\textbf{\citep{Xia2024}} & 2024 & 3D-TP \& ATC & RL  PID Controller & Avoidance of obstacles \\
\textbf{\citep{Wei2023Joint}} & 2023 & 2D-TP \& TA & RL (Deep Q-Network) & Min. energy consumption and latency \\
% \textbf{\citep{Almasaeid2025comcom}} & 2025 & 2D Position \& TA & RL (Greedy Approximation) & Min. energy consumption rate \\
\textbf{\citep{Cui2020}} & 2020 & 3D-TP \& TA & MARL & Max. long-term resource efficiency \\
\revise{\textbf{\citep{Zhou2022two}}} & \revise{2022} & \revise{3D-TP \& TA}  & \revise{Successive convex approximation} & \revise{Min. latency} \\
\revise{\textbf{\citep{Huang2023online}}} & \revise{2023} & \revise{3D-TP \& TA}  & \revise{DRL} & \revise{Min. latency} \\
\textbf{\citep{Wei2019}} & 2019 & 2D-TP \& TA \& RA & GA \& Stepwise Approximation & Min. energy consumption and latency \\
\textbf{\citep{li2020energy}} & 2020 & 3D-TP \& TA \& RA & SCA \& ADMM (Alternating Direction Method of Multipliers) & Max. UAV energy efficiency \\
\textbf{\citep{Qin2023Multi}} & 2023 & 3D-TP \& TA \& RA & RL (MADDPG) & Min. energy consumption and latency \\
\textbf{\citep{Tan2024TD3}} & 2024 & 3D-TP \& TA \& RA & Differential Evolutionary (DE) \& RL (TD3) & Optimal trajectory and convergence \\
\textbf{Our Work} & 2025 & 3D-TP \& TA \& RA \& ATC & ACS-DS \& PSO \& FEAR-PID & Min. energy consumption and latency \\  
\midrule
\multicolumn{5}{l}{\textit{\textbf{Note:} TP = Trajectory Planning, TA = Task Assignment, RA = Resource Allocation, ATC = Attitude Control}} \\
\bottomrule
\label{tab:related}
\end{tabular}
}
\end{table*}

Our work in this paper combines the techniques mentioned above that have been demonstrated to be effective in respective processes, including FEAR-PID attitude control, ACS-based trajectory planning, and PSO-based task allocation and resource allocation. Following this foundation, we further improve several aspects of the entire framework, such as 1) the hardware of the quadrotor is designed to match the efficient performance of the FEAR-PID control system, 2) two enhanced anti-lockout mechanisms (decoupling and safety values) in the classical ACS for trajectory planning, and 3) a modified PSO approach with improved efficiency that is more appropriate for task assignment and resource allocation in large-scale UAV-assisted fog computing systems.

%%%%%%%%%%%%%%%%%%%%%%%%%%%%%%%%%%%%%%%%%%%
%\vspace{-0.1cm} % *****
\section{System Model} \label{subsec: Network}
%\vspace{-0.1cm} % *****
\subsection{Network Structure and Components}
\begin{figure}[htbp]
\centering
\includegraphics[width=0.95\linewidth]{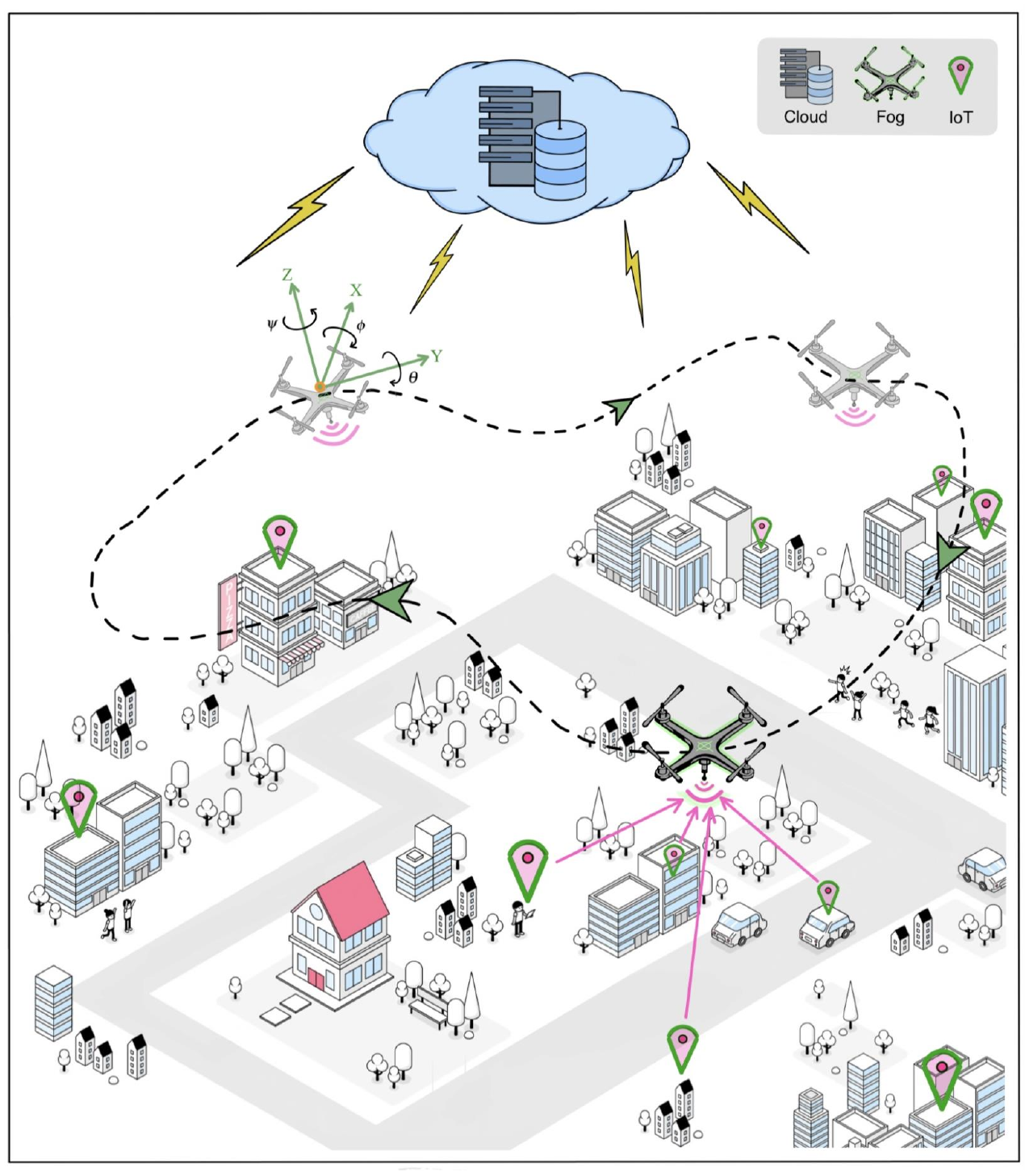}
\caption{The structure of UAV-assisted fog computing network.}
\label{fig: Network}
\end{figure}

We consider a UAV-assisted fog computing system in a three-dimensional Euclidean space. The system consists of a quadrotor UAV, a remote data center (DC) in the cloud, and $K$ mobile IoT devices (MDs). A demonstration of the key structures in the system is shown in Fig.~\ref{fig: Network}. 

\vspace{0.3cm}

The positions of $K$ MDs are randomly distributed according to a Poisson Point Process (PPP), with the coordinate of the $j$th MD denoted by $\boldsymbol{M}_j=\left[x_j, y_j,z_j \right]$. For convenience, we discretize a continuous time horizon with length $T$ uniformly into $N$ timeslots, and thus each timeslot has a length of $L=\frac{T}{N}$. We assume that $N$ is sufficiently large, or equivalently, the timeslots are sufficiently short, such that the position of the UAV can be considered to be fixed within each timeslot. \revise{Note that this timeslot refers to the high-level system decision interval for task allocation, trajectory updates, and communication resource allocation, rather than the low-level attitude control loop of the UAV. The attitude control loop typically operates at 100--500 Hz and is handled automatically by the flight controller, whereas the system-level decision timeslot in UAV or general mobile edge computing (MEC) systems commonly ranges from 0.1 s to several seconds and has no direct correspondence with the flight control loop frequency.} We use $[x(t), y(t), z(t)]$, where $t \in \{1,2, \cdots, N\}$, to denote the UAV position at the $t$th timeslot. The UAV is powered by a battery with a maximum capacity of $B_c$. The height and speed are restricted to not exceed $z^{\max}$ and $v^{\max}$, respectively, at all times. 

We consider that task arrivals from each MD conform to a Poisson process, with an arrival rate $\lambda_j$ from the $j$th MD.  We use $s_{i j}(t)$, which is assumed to be exponentially distributed, to denote the data size of the $i$th task from the $j$th MD to be processed at the \revise{$t$th} timeslot, and $c_{i j}$ to denote the number of CPU cycles per bit required to process the task. 
As we introduced in~\citep{Liu2022}, the proportion of channels allocated to the $j$th MD would be based on the Gamma distribution, %that is,

\begin{equation}\label{eqn: channel}
%\vspace{-0.05cm} % *****
P_j(t)= \frac{\beta^{\alpha}}{(\alpha-1)!}s_{ij}(t)^{\alpha-1}e^{-\beta s_{ij}(t)},
\end{equation}
where $\alpha$ and $\beta$ are the shape and rate parameters in the Gamma distribution. For demonstration purposes, we set $\alpha=\beta=2$ throughout the paper. 

Given the total number of available channels $N_c$, the number of channels allocated to the $j$th MD at the $t$th timeslot is $C_j(t) =\left[ N_c \cdot P_j(t)\right]$ where $\left[x\right]$ rounds $x$ to the nearest integer. The rationale of this arrangement is to prevent the channel from being monopolized by extremely large tasks, and thus to ensure a certain level of transmission efficiency for all tasks.

For the $i$th task from the $j$th MD, we denote the energy consumption and delay as $E_{ij}$ and $D_{ij}$, respectively. For each task, the quadrotor UAV needs to move sufficiently close to the MD that initiates the task through a planned trajectory, in order to receive and further process the task. As mentioned earlier, a task may be executed locally, in the fog layer by the UAV, or further offloaded to the data centers in the central cloud. The computing results of any offloaded task need to be transmitted back to the initiating MD. In this context, we define $\mathbf{o}_{ij}(t)=(o_{ij}^{\text{MD}}(t), o_{ij}^{\text{UAV}}(t), o_{ij}^{\text{DC}}(t))$ as an array consisting of binary variables that indicate the specific execution location of the $i$th task from the $j$th MD during the $t$th timeslot. 

For ease of reference, \revisenew{Table~\ref{tab: notation} summarizes the core notations used throughout the system model in Section~\ref{subsec: Network}; experiment-specific parameter values will be provided separately in Tables~\ref{tab:MODEL PARAMETER SETTINGS} and~\ref{tab:ALGORITHM PARAMETER SETTINGS} in the results section.}

\begin{table*}[t]
\centering
\caption{KEY NOTATIONS IN THE SYSTEM MODEL}
\label{tab: notation}
\resizebox{\textwidth}{!}{ 
\renewcommand{\arraystretch}{1}
\begin{tabular}{>{\centering}m{2.5cm} m{7cm} >{\centering}m{2.5cm} m{7cm}}
\toprule 
\textbf{Notation} & \textbf{Definition} & \textbf{Notation} & \textbf{Definition} \\
\midrule
\multicolumn{4}{c}{\revisenew{\textbf{Network, Task Assignment, and Resource Allocation-Related Parameters}}} \\
\hline
$K$ & The total number of MDs & $L$ & The length of every timeslot \\
\hline
$T$ & The total number of timeslots required to complete all tasks & $(x_j,y_j,z_j)$ & The location of the $j$th MD \\
\hline
$(x(t),y(t),z(t))$ & The location of the UAV at the $t$th timeslot & $\mathbf{v}(t)$ & The UAV's velocity vector at the $t$th timeslot \\
\hline
$d_j(t)$ & The distance between the UAV and $j$th MD at the $t$th timeslot & $B_c$ & The battery capacity of the UAV \\
\hline
$N_j$ & The total number of tasks from the $j$th MD & $N_c$ & The total number of wireless channels \\
\hline
$C_j(t)$ & The number of channels assigned to the $j$th MD at the $t$th timeslot & $N_{b}$ & The number of UAV propeller blades \\
\hline
$\lambda_j$ & \revisenew{Task arrival rate of the $j$th MD} & $s_{ij}(t)$ & \revisenew{Input data size of the $i$th task from the $j$th MD at timeslot $t$} \\
\hline
$c_{ij}$ & \revisenew{CPU cycles per bit required for the $i$th task from the $j$th MD} & $P_j(t)$ & \revisenew{Channel allocation proportion for the $j$th MD (Gamma-based)} \\
\hline
$\alpha,\beta$ & \revisenew{Shape and rate parameters in the Gamma-based channel allocation} & $\mathbf{o}_{ij}(t)$ & \revisenew{Task execution decision vector (MD/UAV/DC) for task $(i,j)$ at timeslot $t$} \\
\hline
\multicolumn{4}{c}{\revisenew{\textbf{UAV Dynamics and Propulsion Parameters}}} \\
\hline
$m$ & \revisenew{UAV mass} & $g$ & \revisenew{Gravitational acceleration} \\
\hline
$I_x,I_y,I_z$ & \revisenew{Rotational inertia around $x,y,z$ axes} & $l$ & \revisenew{Distance from UAV center to rotor} \\
\hline
$\mathbf{U}$ & \revisenew{Dynamics input vector $(U_1,U_2,U_3,U_4)^{T}$} & $\omega_i(t)$ & \revisenew{Angular velocity of rotor $i$ at timeslot $t$} \\
\hline
$R$ & The radius of UAV propeller blades & $\gamma$ & The mounting angle of UAV propeller blades \\
\hline
$P_{w}, P_{t}$ & Width and thickness of UAV propeller blades & $p_u(t), p_d(t), p_j(t)$ & The transmission power of the UAV, the DC, or the $j$th MD at the $t$th timeslot \\
\hline
$C_T, C_M$ & \revisenew{Thrust and torque coefficients of the propellers} & $Q_{\gamma}(t)$ & \revisenew{Yaw-/mounting-angle-dependent coefficient in the propeller model} \\
\hline
\multicolumn{4}{c}{\revisenew{\textbf{Computation and Queuing Parameters}}} \\
\hline
$f(t), f_{j}(t)$ & The processing frequency of equipment for UAV, or the $j$th MD at the $t$th timeslot & \revise{$\delta_u, \delta_j$} & \revise{Computation-energy coefficients for UAV and MD} \\
\hline
\revise{$\bar{Q}$} & \revise{UAV computation queue-capacity parameter (bit-equivalent)} & $\mathbb{S}(t)$ & The weighted sum of network energy consumption and delay at the $t$th timeslot \\
\bottomrule
\end{tabular}
} 
\end{table*}

%%%%%%%%%%%%%%%%%%%%%%%%%%%%%%%%%%%%%%%%%%%
\subsection{The Controllable Structure of a Quadrotor UAV} \label{sec:structure}

Quadrotor UAVs are capable of achieving six degrees of freedom in pitch, yaw, roll, vertical, fore and aft, and lateral motion by translating and rotating along the $x$, $y$ and $z$ axes. The process is accomplished by controlling four kinetic input quantities of the motor, denoted by $\mathbf{U}=\left(U_{1}, U_{2}, U_{3}, U_{4}\right)^{T}$,  and six output control variables that include the changes of movements $\mathbf{v}=(\Delta x, \Delta y, \Delta z)$ on the $x$, $y$, and $z$ directions, and the changes of attitude angles $\mathbf{A}=(\Delta \phi, \Delta \theta, \Delta \psi)$ resulting from rotation on the three axes. 

Specifically, $\mathbf{v}$ stands for linear velocity, which is the speed at which the UAV moves in real space and determines the location and speed of the UAV in the map. $\mathbf{A}$ determines angular velocities, the attitude change and the turning speed of the UAV.

According to the Newton-Euler theorem, the nonlinear dynamics equations of a quadrotor UAV are~\citep{Huang2019},

\begin{equation} \label{equ: dynamics_equation}
\resizebox{0.8\linewidth}{!}{
    $
    \left\{
    \begin{array}{l}
    \ddot{x}(t) = \frac{1}{m} U_{1}(t)(\sin \phi \sin \psi+\cos \phi \sin \theta \cos \psi) \\ 
    \ddot{y}(t) = \frac{1}{m} U_{1}(t)(-\sin \phi \sin \psi+\cos \phi \sin \theta \sin \psi) \\ 
    \ddot{z}(t) = \frac{1}{m} U_{1}(t)(\cos \phi \cos \theta)-g \\ 
    \ddot{\phi}(t) = \frac{1}{I_{x}}\left(l U_{2}(t)-\dot{\theta} \dot{\psi}\left(I_{z}-I_{y}\right)\right) \\ 
    \ddot{\theta}(t) = \frac{1}{I_{y}}\left(l U_{3}(t)-\dot{\phi} \dot{\psi}\left(I_{x}-I_{z}\right)\right) \\ 
    \ddot{\psi}(t) = \frac{1}{I_{z}}\left(U_{4}(t)-\dot{\phi} \dot{\theta}\left(I_{y}-I_{x}\right)\right) 
    \end{array}
    \right.
    $
}
\end{equation}
where the left hand side of each equation represents the second order derivative of the relevant variable with respect to $t$. $I_{x}$, $I_{y}$, and $I_{z}$ are the rotational inertia of the UAV body around its own $x$, $y$, and $z$ axes, respectively. $I_{x}$ and $I_{y}$ are approximated based on the assumption of structural symmetry of the quadrotor UAV. Meanwhile, $l$ is the distance from the center of mass of the body to the center of the rotor, which is also equal to half of the airframe wheelbase.

\begin{figure}[ht]
\centering
\includegraphics[width=0.95\linewidth]{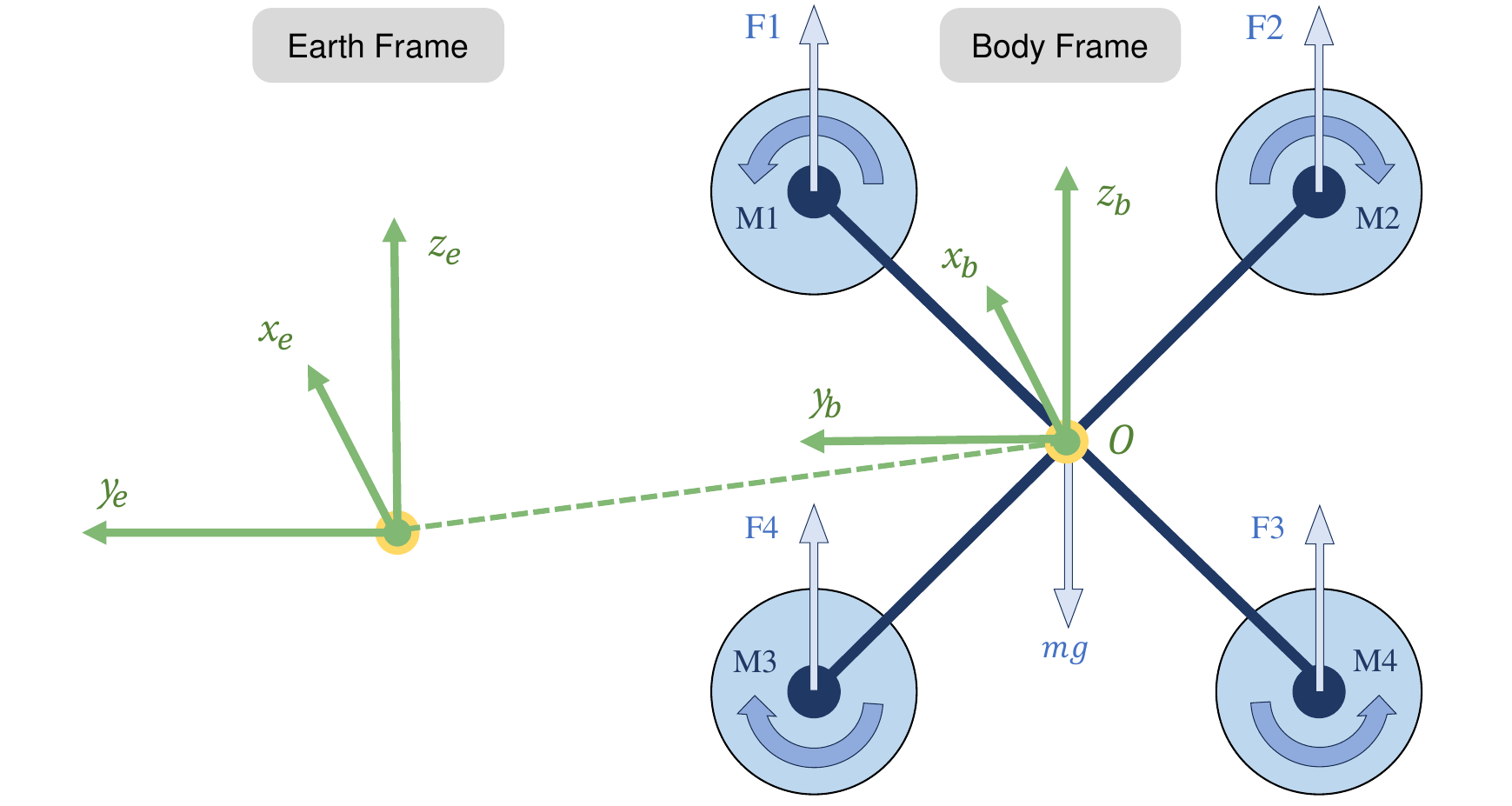} 
\caption{Schematic diagram of a quadrotor UAV in motion, \revisenew{where subscripts $e$ and $b$ denote the earth and body frames, respectively, and $(x_e,y_e,z_e)$ and $(x_b,y_b,z_b)$ denote their coordinate axes associated with the translational directions $x,y,z$.}}
\label{fig: coordinate}
\end{figure}

Fig.~\ref{fig: coordinate} shows the schematic diagram of a quadrotor UAV in motion. Two pairs of motors $(M1, M4)$ and $(M2, M3)$ rotating in opposite directions are used to eliminate counter-torque. Moreover, $m$ is the mass of the UAV. $g$ is the acceleration of gravity, $I_{x}$, $I_{y}$, and $I_{z}$ are the rotational inertia of the UAV body around $x$, $y$, and $z$ axes, respectively. \revisenew{In Fig.~\ref{fig: coordinate}, subscripts $e$ and $b$ denote the earth (inertial) frame and body frame, respectively; the corresponding coordinate axes are denoted as $(x_e,y_e,z_e)$ and $(x_b,y_b,z_b)$. These axes correspond to the translational directions associated with the variables $x,y,z$ in~\eqref{equ: dynamics_equation}.}

The input and output quantities interact with each other to change the UAV motion by adjusting the rotational speed of the four rotors (propellers). The thrusts $F_{i}(t),(i=1,2,3,4)$ provided by the motors are directly proportional to the square of the angular velocity of the rotors (propellers) of the corresponding motors $\omega_{i}(t),(i=1,2,3,4)$, namely,
\begin{equation}
F_{i}(t)=C_{T}(t) \omega_{i}(t)^{2}, \label{eqn:fi}
\end{equation}
where $C_{T}(t)$ is the thrust coefficient of the propellers. The torque representing the magnitude of the torque to overcome the air resistance is given by 
\begin{equation}
M_{p}(t)= C_{M}(t) \omega_{i}(t)^{2}, 
\end{equation}
with $C_{M}(t)$ as the torque coefficient. %Both $C_{T}(t)$ and $C_{M}(t)$ are determined by the parameters of the quadrotor UAV as well as the air density, details of which are set out in the next subsection. 

For more details on principles related to attitude control and the calculation of the dynamics input $\mathbf{U}$, see \ref{appendixA}.

%%%%%%%%%%%%%%%%%%%%%%%%%%%%%%%%%%%%%%%%

%%%%%%%%%%%%%%%%%%%%%%%%%%%%%%%%%%%%%%%%%%%%%%%%%%%%%%%%%%%%%%%%%%%
\subsection{Rotor Propeller Design of Quadrotor UAV} \label{subsec: Rotor propeller} 
\begin{figure}[ht]
\centering
\includegraphics[width=0.95\linewidth]{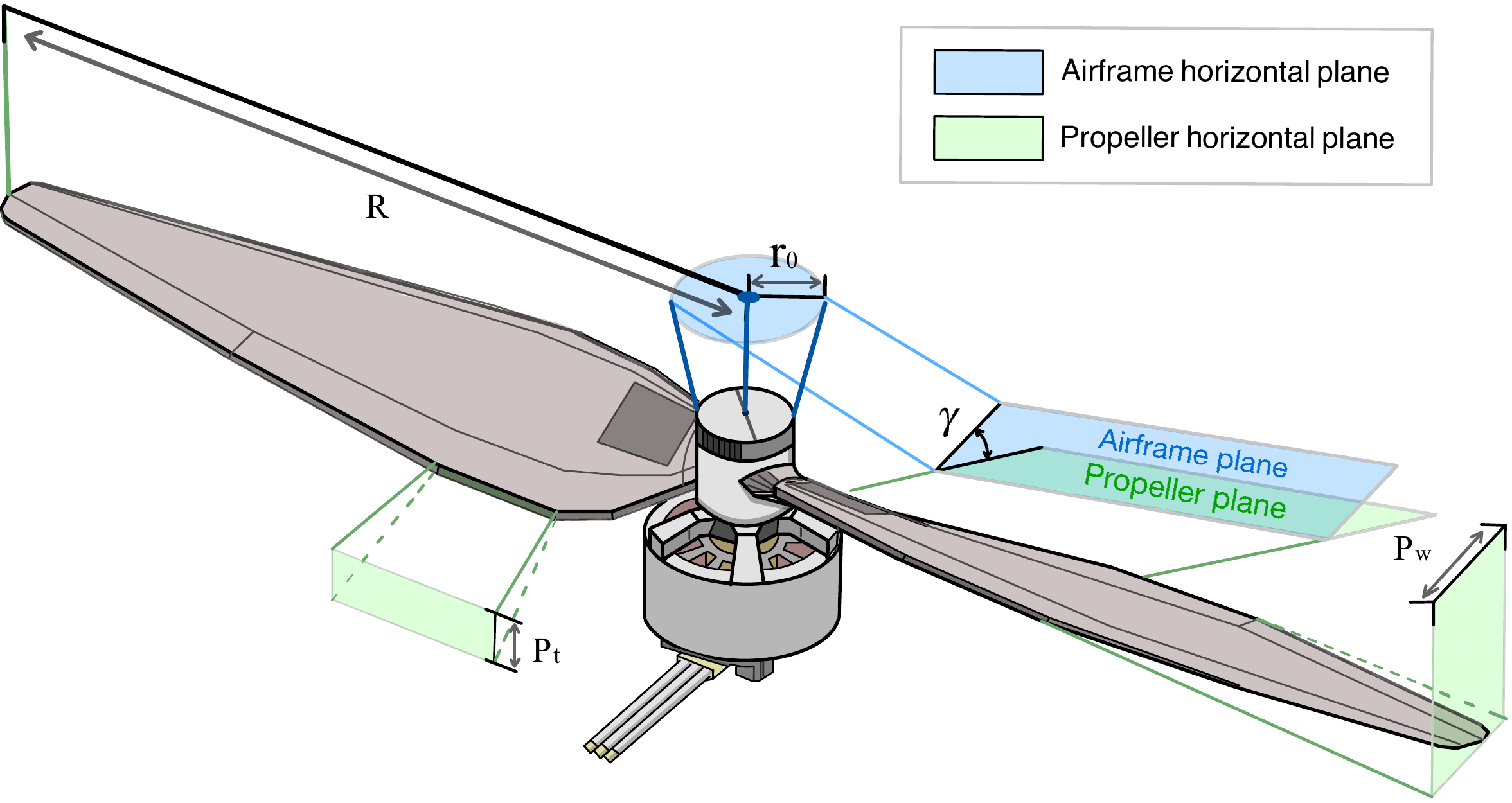}
\caption{Schematic of quadrotor UAV propeller hardware.}
\label{fig: propeller}
\end{figure}

\revise{The design and performance modeling of multirotor UAVs, including propeller configurations and power systems, have been extensively studied~\citep{Biczyski2020,Zhu2021}.} We now introduce the factors that would affect $C_{T}(t)$ and $C_{M}(t)$. The propeller is the power source of the UAV. Fig.~\ref{fig: propeller} shows the structure of the quadrotor UAV propeller, where the propeller radius $R$ refers to the sum of the blade length $r$ and the radius of the mounting interface $r_0$. The blade mounting angle is the angle between the propeller blades and the plane of the UAV airframe, denoted by $\gamma$. Along with the blade width $P_w$, thickness $P_t$, we define the torque coefficient as,
\begin{equation}
C_M(t) = 2R \cdot C_T(t)=\frac{\left(\frac{1}{2 \pi}\right)^2 \rho_a(t)(2 R)^5}{N_B \int_{r_0}^R P_t^4 P_w Q_{\gamma}(t) dr}, 
\end{equation}
where $N_B$ is the number of propeller blades, and $Q_{\gamma}(t)$ is determined by the propeller blade mounting angle and yaw angle in the current state. \revise{The power loading and disk loading characteristics of propellers significantly influence the overall energy efficiency~\citep{Pollet2020}.} $\rho_{a}(t)$ is the air density determined by the atmospheric pressure at the location of the UAV and the absolute temperature, specifically,

\begin{equation}
\rho_{a}(t) = \frac{P_{g} \cdot M_{g}} {R_{g} \cdot \Theta(t)},
\end{equation}
where $P_{g}$ is the standard atmosphere, $M_{g}$ is the molecular weight of gas, $R_{g}$ is the gas constant, $\Theta(t)$ is the outside ambient temperature, $z(t)$ is the height of flight at the location of the UAV, and $M_{g}$ and $R_{g}$ are constants in the ideal gas state.

\begin{figure}[ht]
\centering
\includegraphics[width=0.8\linewidth]{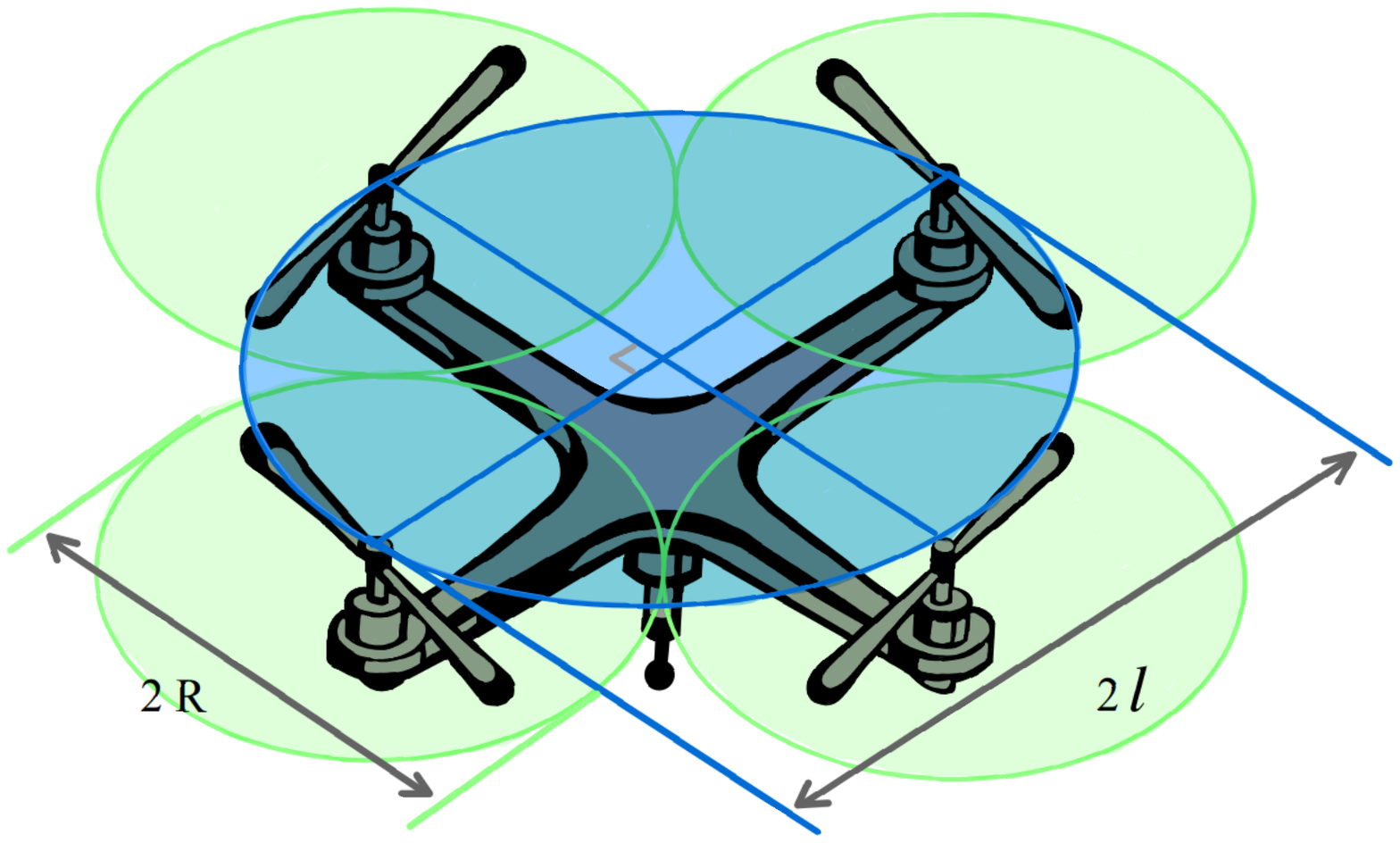} 
\caption{Propeller radius in relationship to wheelbase for quadrotor UAV.}
%\vspace{-0.25cm} % *****
\label{fig: body}
\end{figure}

The values of propeller radius $R$ and airframe wheelbase $2l$ need to be set properly, in order to avoid collisions between neighboring propellers and underpowered situations. We illustrate the structural relationship in Fig.~\ref{fig: body}. In a symmetrically structured quadrotor configuration, the diagonal separation distance between neighboring propellers is $\sqrt{2} l$, and the total width of the two propellers ($2R$) must be less than this distance to avoid collision, namely $2 R<\sqrt{2} l$. On the other hand, a too small $R$ (less than $l/3$) will not provide enough lift for the UAV according to~\eqref{eqn:fi}. Thus, there is a constrained relationship between $R$ and $l$, namely $\frac{\sqrt{2}}{2}l \geqslant R \geqslant \frac{1}{3}l$.

%%%%%%%%%%%%%%%%%%%%%%%%%%%%%%%%%%%%%%%%%%%%%
\subsection{Transmission Model} \label{subsec: Transmission}
%\highlight{A paragraph may be needed here to summarize the variables that appear, such as $e_{i j}^{\text{MU}}(t)$, $e_{i j}^{\text{UM}}(t)$, $e_{i j}^{\text{UD}}(t)$ and $e_{i j}^{\text{DU}}(t)$. Includes $\xi_{j, u}(t)$, $\xi_{u, j}(t)$, $\xi_{u, d}(t)$ and $\xi_{d, u}(t)$. Includes $r_{j, u}(t)$, $r_{u, j}(t)$, $r_{u, d}(t)$ and $r_{d, u}(t)$.}

We define a tuple $(m,n)$ to represent a transmission from $m$ to $n$, where $m$ and $n$ can be $j \in \{1,2,\cdots, K\}$ (representing the $j$th MD), $u$ (the UAV), or $D$ (the DC).

We consider $\zeta_j$ to be a parameter determined by the carrier frequency $f_c$ and obstacle density,  
corresponding to the proportion of the Line of Sight (LoS) path obstructed between the $j$th MD and the UAV, while the elevation angle of the link is $\theta_{j}$. 

\revise{In our deployment, the communication antenna of the UAV is bottom-mounted, and the MDs are always located on the ground surface. Therefore, the UAV always maintains a higher altitude than all MDs, ensuring that no geometric airframe shadowing occurs. The aggregated blockage effects—including terrain and man-made obstacles—are modeled through a parameter $\zeta_j$, as defined in the following equation.}

The probability to establish a LoS connection between the $j$th MD and the UAV is,
\begin{equation}
P^{\text{LoS}}_j(t)=\frac{1}{1+\zeta_j \cdot \exp \left(-\left(\theta_{j}-\zeta_j  \right)\right)}.
\end{equation}

The Signal-to-Interference-plus-Noise Ratio (SINR) for $(m,n)$ at the $t$th timeslot is,
%is one critical factor affecting the quality of wireless communication, and from the Doppler effect of Slow fading, the SINR 
\begin{equation}
\xi_{(m,n)}(t)=\frac{p_m(t)}{\mathcal{L}_{(m,n)}(t) \cdot \left(I_{(m,n)}(t) + \sigma_{T}^2\right)},
\end{equation}
where $p_m(t)$ is the transmit power of node $m$ at the $t$th timeslot. We denote by $P^{\text{LoS}}_{(m,n)}(t)$ the LoS probability of link $(m,n)$, which equals $P^{\text{LoS}}_j(t)$ for MD--UAV links and is set to $1$ for the UAV--DC backhaul link for simplicity. 
$\mathcal{L}_{(m, n)}(t)= P^{\text{LoS}}_{(m,n)}(t) \cdot \left(\frac{4 \pi f_c}{c} d_{(m,n)}(t)\right)$ is the path loss, $c$ and $d_{(m,n)}(t)$ stand for the speed of light and the distance between $m$ and $n$, respectively. \revise{Here, $I_{(m,n)}(t)$ denotes the aggregate external interference from concurrent transmissions outside our UAV-assisted fog network, which we model as additive noise for simplicity.} $\sigma_{T}^2$ is the thermal noise. \revise{Our transmission model assumes orthogonal channel allocation within the UAV-assisted network (via the channel allocation mechanism in Section~3.1), which ensures negligible co-channel interference among the $K$ MDs communicating with the UAV. This simplified interference modeling approach is standard in MEC and fog computing literature, where detailed multi-cell interference coordination would require additional coordination mechanisms and is considered as future work.}

We can then obtain the effective transmission rate for $(m,n)$ at $t$th timeslot as,
\begin{equation}
r_{(m, n)}(t)=W_{(m, n)} \log _2\left(1+\xi_{(m, n)}(t)\right),
\end{equation}
where $W_{(m, n)}$ is the bandwidth of the wireless link. 

%%%%%%%%%%%%%%%%%%%%%%%%%%%%%%%%%%%%%%%%%%%%
\subsection{Energy Consumption Model} \label{subsec: energy}
We consider that the total energy consumption includes the consumption for the transmission and computation during the task offloading process, as well as those for the movement of the UAV. Hereafter, we use the binary variables $o_{ij}^{\text{MD}}(t)$, $o_{ij}^{\text{UAV}}(t)$, and $o_{ij}^{\text{DC}}(t)$ to specify the assignment of the $i$th task from the $j$th MD at the $t$th timeslot. A value of $1$ indicates that the task is assigned to the corresponding location.

\subsubsection{UAV movement} We consider that the motion state of the UAV is directly related to the rotational speed of the four DC motors. \revise{The energy consumption for UAV movement is highly dependent on the payload and flight characteristics~\citep{Karam2024,Silva2024,Zhu2021}.} 
Thus, the energy consumption due to UAV movement at the $t$th timeslot is,
\begin{equation}
E^{\mathrm{MOV}}(t)=\sum_{i=1}^{4}\left(I_{i}(t)\cdot L\right)=\sum_{i=1}^{4}\left(\frac{V_{m}-\omega_{i}(t) P_{m}}{R_m} \cdot L\right),
\label{eqn: move_enegy}
\end{equation}
where $I_{i}(t)$ is the motor current, which can be calculated by \eqref{eqn: control_quantity} and \eqref{equ: dynamics_equation} given that $U$ and $\omega_{i}(t) (i=1,2,3,4)$ (the linear speed of $i$th motor rotation (r/min) at the $t$th timeslot) are obtained by our proposed attitude control method. $V_{m}$ is the motor rated voltage, $R_{m}$ is the rated resistance of the UAV motor circuit, and $P_{m}$ is the electric potential constant determined by the motor structure.

%\subsubsection{Unstable deflection}
%The larger the attitude angle is during flight, the more unstable the UAV's flight state is and the less smooth the flight trajectory is. Therefore, we consider adding a smoothing cost to increase the stability of the UAV trajectory.

%\begin{equation}
%E_{u}^{3}(t)=\sum_{i=2}^{N}\left|\mathbf{A}_{t}-%\mathbf{A}_{t-1}\right|.
%\end{equation}

%Combining the above three points, the energy consumption of the UAV for moving and smoothing the airframe in the auxiliary calculation process is $E_{u}(t)=\sum_{i=1}^{3}(E_{u}^{i}(t))$. \timothy{$\mathbf{A}$ was defined in the previous section, but it is better to mention the definition again here, especially now that you have $\mathbf{A}_t$. Also, I didn't see the definition of $E_{u}^{1}(t)$.}

%%%%%%%%%%%%%%%%%%%%%%%%%%%%%%%%%%%%%%%%%%%%%%%%%%%%%%%%%%%%%%%%%%%%%%%%%%

\subsubsection{Energy consumption for transmission} If a task is assigned to the UAV, two segments of transmission will be incurred, namely from the MD to the UAV and the UAV back to the MD. On the other hand, if a task is assigned to the DC, two additional segments, namely from the UAV to the DC and from the DC back to the UAV, will be incurred. Therefore, energy consumption at the $t$th timeslot for transmitting the $i$th task from the $j$th MD is,
\begin{equation}\label{eqn: Energy_transmission}
\resizebox{0.9\linewidth}{!}{
    $
    \begin{aligned}
    e_{i j}^{\text{TR}}(t) =& \left(o_{ij}^{\text{UAV}}(t) + o_{ij}^{\text{DC}}(t)\right) 
    \left(\frac{p_{j}(t) s_{i j}(t) }{C_j(t) r_{(j, u)}(t)} + \frac{p_{u}(t) s_{i j}(t) }{C_j(t) r_{(u, j)}(t)}\right) \\
    &+ o_{ij}^{\text{DC}}(t) 
    \left(\frac{p_{u}(t) s_{i j}(t) }{C_j(t) r_{(u, D)}(t)} + \frac{p_{D}(t) s_{i j}(t) }{C_j(t) r_{(D, u)}(t)}\right).
    \end{aligned}
    $
}
\end{equation}
where $C_j(t)$ is the number of channels assigned to the $j$th MD. 

\subsubsection{Energy consumption for computation}
We consider that the additional energy consumption for computing a finite number of tasks in the DC is negligible as the DC is assumed to be always active in processing tasks from different sources. Therefore, the energy consumption for computing $i$th task of the $j$th MD can be obtained by
\begin{equation}\label{eqn: Energy_comp}
\resizebox{\linewidth}{!}{
$
e_{ij}^{\mathrm{COMP}}(t)=o_{ij}^{\text{UAV}}(t)\delta_{u}s_{i j}(t) c_{i j}\left(f(t)\right)^{2}+o_{ij}^{\text{MD}}(t)\delta_{j}s_{i j}(t) c_{i j}\left(f_j(t)\right)^{2},
$
}
\end{equation}
% \begin{equation}\label{eqn: Energy_MD}
% e_{ij}^{\mathrm{MD}}(t)=\delta_{j} \cdot o_{ij}(t)^{\text{MD}} \cdot s_{i j} c_{i j}\left(f_{j}(t)\right)^{2},
% \end{equation}
\revise{where $\delta_{u}$ and $\delta_{j}$ are the computation-energy coefficients (unit: J/(cycle$^2\cdot$bit)) for the UAV and the $j$th MD, respectively. Specifically, $\delta_u = 1.2\times10^{-28}$ J/(cycle$^2\cdot$bit) for the UAV and $\delta_j = 3.0\times10^{-28}$ J/(cycle$^2\cdot$bit) for MDs, where the higher value for MDs reflects their weaker hardware efficiency.}

%%%%%%%%%%%%%%%%%%%%%%%%%%%%%%%%%%%%%%%%%%%%%%%%%%%%%%%%%%%%%%%%%%%%%%%%%%
\subsubsection{Overall Energy Consumption} 
\revise{Based} on the discussions above, the overall energy consumption at the $t$th timeslot can be expressed as,  
\begin{equation} \label{eqn:total_energy}
\mathbb{E}(t)=\sum_{j=1}^{K} \sum_{i=1}^{N_j} \left[e^{\mathrm{TR}}_{i j}(t)+e^{\mathrm{COM}}_{i j}(t)\right] + E^{\mathrm{MOV}}(t),
\end{equation} 
where $N_j$ is the total number of tasks the $j$th \revise{MD}.

%%%%%%%%%%%%%%%%%%%%%%%%%%%%%%%%%%%%%%%%%%%%%%%%%%%%%%%%%%%%%%%%%%%%
\subsection{Delay Model}
%\highlight{A paragraph may be needed here to summarize the variables that appear, such as $d_{i j}^{\text{MU}}(t)$, $d_{i j}^{\text{UM}}(t)$, $d_{i j}^{\text{UD}}(t)$ and $d_{i j}^{\text{DU}}(t)$.}

We consider three delay components: transmission, computation, and queuing. %The efficiency of the CPU in the computational element, i.e., the Virtual Machine (VM), on which the UAV is equipped determines the task processing capability of the UAV, while factors such as the task processing efficiency and the task arrival rate together determine the task completion time and the queuing time in case of insufficient computational space. When the task processing rate is less than the arrival rate the computational space is insufficient and the queuing of tasks during transmission and computation needs to be considered.

\subsubsection{Transmission delay}
The transmission delay for the $i$th task from the $j$th MD at the $t$th timeslot is, %\timothy{similar to the transmission energy, i revised this equation.}
\begin{equation}\label{eqn: delay_transmission}
\resizebox{\linewidth}{!}{
    $
    \begin{aligned}
    d_{i j}^{\text{TR}}(t) &= o_{ij}^{\text{DC}}(t) \left(\frac{s_{i j}(t)}{C_j(t) r_{(u, D)}(t)} + \frac{s_{i j}(t)}{C_j(t) r_{(D, u)}(t)}\right) \\
    &\quad + \left(o_{ij}^{\text{UAV}}(t) + o_{ij}^{\text{DC}}(t)\right) \left(\frac{s_{i j}(t)}{C_j(t) r_{(j, u)}(t)} + \frac{s_{i j}(t)}{C_j(t) r_{(u, j)}(t)}\right). 
    \end{aligned}
    $
}
\end{equation}
% \begin{equation}\label{eqn: Delay_transmission}
% d_{i j}^{MU}(t)=\frac{s_{i j} \cdot o_{ij}^{\text{MD}}}{C_j(t) r_{j, u}(t)}.
% \end{equation}
% Similarly, we can define the delay of task transmission for the processes from the UAV to the $j$th MD as $d_{i j}^{UM}(t)=\frac{s_{i j} \cdot o_{ij}^{\text{MD}}}{C_j(t) r_{u, j}(t)}.$, from the UAV to the DC as $d_{i j}^{UD}(t)=\frac{s_{i j} \cdot o_{ij}^{\text{DC}}}{C_j(t) r_{u, d}(t)}$, and from the DC to the UAV as $d_{i j}^{DU}(t)=\frac{s_{i j} \cdot o_{ij}^{\text{DC}}}{C_j(t) r_{d, u}(t)}$.
% }

\subsubsection{Computation delay}
The delay for computing the $i$th task from the $j$th MD can be expressed as,
\begin{equation}\label{eqn: delay_comp}
\resizebox{0.78\linewidth}{!}{
$
d_{ij}^{\mathrm{COMP}}(t)=o_{ij}^{\text{UAV}}(t)\frac{s_{i j}(t) c_{i j}}{f(t)}+o_{ij}^{\text{MD}}(t)\frac{s_{i j}(t) c_{i j}}{f_j(t)}.
$
}
\end{equation}
\revise{where $f(t)$ (unit: GHz) denotes the UAV's processing frequency, bounded by $f^{\min} = 1.0$ GHz and $f^{\max} = 3.0$ GHz, representing a high-performance onboard processor. Similarly, $f_j(t)$ (unit: GHz) denotes the $j$th MD's processing frequency, bounded by $f_j^{\min} = 0.5$ GHz and $f_j^{\max} = 2.0$ GHz. The UAV distributes its processing frequency $f(t)$ among all assigned tasks according to the joint optimization algorithm, which minimizes the overall weighted energy--delay cost. Thus, resource allocation is implicitly determined by the solution of the optimization problem rather than by a fixed scheduling rule.}

% \begin{equation}
% d_{ij}^{\mathrm{MD}}(t)=\frac{o_{ij}^{\text{MD}}(t) \cdot s_{i j} c_{i j}}{f_j(t)},
% \end{equation}

% \begin{equation}
% d_{ij}^{\mathrm{UAV}}(t)=\frac{o_{ij}^{\text{UAV}}(t) \cdot s_{i j} c_{i j}}{f(t)}.
% \end{equation}

Note that as the computing resources at the DC are sufficient for all practical purposes, we assume that the computation delay at the DC is negligible.

\subsubsection{Queuing delay} 
\revise{We consider two types of queuing delays: transmission queuing and computation queuing. Transmission queuing occurs when the arrival rate exceeds the communication service rate, while computation queuing occurs when the aggregated task workload exceeds the UAV's effective computing rate.}

\revise{Our queuing model adopts an M/M/1-type abstraction, which is a standard first-order modeling approach in MEC literature for capturing queuing behavior under stochastic task arrivals~\revisenew{\citep{zukerman2013introduction,li2022data}}. This abstraction enables tractable analysis while providing reasonable approximations of system performance under moderate to high load conditions. While more detailed queueing network models (e.g., multi-server queues or priority-based scheduling) could offer finer granularity, they would significantly complicate the optimization problem formulation. Our simplified model remains compatible with such extensions, which could be incorporated as future work when detailed scheduling policies are required.}

Consider that the arrival rate for transmission from the $j$th MD to UAV is $\lambda_{j}$, and the processing rate at the UAV for receiving transmissions is $[L \cdot C_j(t) \cdot r_{j, u}(t)]/s_{j}(t)$, where $s_{j}(t) = \sum_{i=1}^{N_j} s_{ij}(t)$ is the sum of tasks data size from that MD. Let $\pi_j$ denote the probability that an arriving task from the $j$th MD does not enter the transmission/computation queue (e.g., executed locally), so the effective arrival rate is $\lambda_{j}(1-\pi_j)$. If the total arrival rate is larger than the processing rate, a queuing delay for transmission will be incurred~\citep{Wei2019}, that is, 
\begin{equation}
\resizebox{0.78\linewidth}{!}{
$
d_{ij}^{\mathrm{Q}}(t)=\frac{\lambda_{j}\left(1-\pi_j\right) s_{j}(t)^{2}}{L \cdot C_j(t) r_{j, u}(t) \cdot \left(L \cdot C_j(t) r_{j, u}(t)-\lambda_{j}\left(1-\pi_j\right) s_{j}(t)\right)}.
$
}
\end{equation}
Similarly, if the total sizes of arriving tasks exceed the computation capacity of the component, queuing delay for computation would be incurred for a proportion of the tasks. The expected queuing delay of the $i$th task from the $j$th MD executed at UAV at $t$th timeslot is,
\begin{equation}\label{eqn: Delay_calculation}
\resizebox{0.78\linewidth}{!}{
$
d_{i j}^{\mathrm{UAVQ}}(t)=\frac{\bar{Q}}{\sum_{j=1}^{K} \lambda_{j}\left(1-\pi_j\right)}-\frac{\sum_{j=1}^{K} \lambda_{j}\left(1-\pi_j\right) s_{j}(t) c_{j}}{\tau f(t) \sum_{j=1}^{K} \lambda_{j}\left(1-\pi_j\right)},
$
}
\end{equation}
\revise{where $\bar{Q} = 5\times10^{7}$ (unit: bit-equivalent) is a queue-capacity parameter representing a queue threshold corresponding to the UAV's onboard computing throughput, and $\tau = 1$ s is a normalization constant that normalizes the aggregated computational workload into an effective service time per task.}

%%%%%%%%%%%%%%%%%%%%%%%%%%%%%%%%%%%%%%%%%%%%%%%%%%%%%%%%%%%%%%%%%%%%%

\subsubsection{Total delay}
The total delay is obtained by summing all delay components mentioned earlier, that is,
\begin{equation} \label{eqn:total_delay}
\mathbb{D}(t)=\sum_{j=1}^{K} \sum_{i=1}^{N_j} \left(d_{i j}^{\mathrm{TR}}(t)+d_{i j}^{\mathrm{COMP}}(t)+d_{i j}^{\mathrm{UAVQ}}(t)+d_{i j}^{\mathrm{Q}}(t)\right).
\end{equation} 
\revise{Congestion in the network manifests through the growth of queue lengths, which directly contributes to the queuing delay components $d_{i j}^{\mathrm{Q}}(t)$ and $d_{i j}^{\mathrm{UAVQ}}(t)$ in the total delay expression. Since our objective function minimizes the weighted sum of delay and energy consumption, congested conditions (characterized by increased queuing delays) are implicitly penalized. This approach eliminates the need for an explicit congestion metric, as the optimization naturally steers the system away from congestion-prone configurations to minimize the overall operational cost.}

\section{Joint Optimization Problem} \label{sec:Joint optimization model}
%\vspace{-0.1cm} % *****

%\vspace{-0.7cm} % *****
We consider that the operational efficiency cost of the network is determined by the weighted sum of the energy and delay components. Therefore, our objective function consists of the energy consumption defined in~\eqref{eqn:total_energy} and the time required to complete all tasks defined in~\eqref{eqn:total_delay}, during the entire period of $T$, as follows.
% \vspace{-0.7cm}
\begin{equation}
\resizebox{0.98\columnwidth}{!}{ 
$
\begin{array}{ll}
\underset{ \{\mathbf{o}_{ij}(t), \mathbf{v}(t), \omega_i(t)|t \in \{0,\cdots,T\}\}}{\operatorname{Min}} & \mathbb{S} = \sum\limits_{t=0}^T[\mathbb{D}(t) + \epsilon \cdot \mathbb{E}(t)] \\
\text {s.t.} 
& \revise{\left(\complement_1\right) \hspace{0.2cm} \omega^{\min} \leq \omega_i(t) \leq \omega^{\max}, \quad \forall i}\\
& \left(\complement_2\right) \hspace{0.2cm} 0 \leq |\mathbf{v}(t)| \leq v^{\max} \\
& \left(\complement_3\right) \hspace{0.2cm} 0 \leq z(t) \leq z^{\max} \\
& \left(\complement_4\right) \hspace{0.2cm} \sum_{j=1}^{K} C_j(t) \leq N_{c}\\
& \left(\complement_5\right) \hspace{0.2cm} f^{\min } \leq f(t) \leq f^{\max } \\
& \left(\complement_6\right) \hspace{0.2cm} f_{j}^{\min } \leq f_{j}(t) \leq f_{j}^{\max } \\
& \left(\complement_7\right) \hspace{0.2cm} p^{\min } \leq p_u(t) \leq p^{\max } \\
& \left(\complement_8\right) \hspace{0.2cm} p_{j}^{\min } \leq p_{j}(t) \leq p_{j}^{\max } \\
& \left(\complement_9\right) \hspace{0.2cm} o_{i j}^{\text{MD}}(t)+o_{ij}^{\text{UAV}}(t)+o_{ij}^{\text{DC}}(t)=1 \\
& \left(\complement_{10}\right) \hspace{0.2cm} o_{i j}^{\text{MD}}(t), o_{ij}^{\text{UAV}}(t), o_{ij}^{\text{DC}}(t) \in\{0,1\} \\
& \left(\complement_{11}\right) \hspace{0.2cm} \sum_{t=0}^{T} \left( E^{\text{UAV}}(t) \right) \leq B_c \\
& i \in\{1, \ldots, N_j\}, j \in\{1, \ldots, K\}
\end{array}\label{eq:problem}
$
}
\end{equation}

% where $r_a^{*}$ and $r_s^{*}$ are task arrival rate and task service rate respectively for each part. To ensure the system is in a steady state, the arrival rate is required to be smaller than the processing rate\citep{Wei2019}. 
The decision variables include the angular velocities $\omega_i(t)$ related to attitude control, the velocity vector $\mathbf{v}(t)$ for trajectory planning, and task assignment decisions $\mathbf{o}_{ij}(t)$. The parameter $\epsilon$ is a weighting factor in the objective function, which flexibly caters to different magnitudes and ranges of the two measurements, as well as %different requirements under various scenarios in UAV-assisted fog computing and 
reflects the preference in terms of the trade-off between the two metrics in various fog computing scenarios. We will demonstrate in the result section that our proposed approach can achieve high performances for a wide range of $\epsilon$ values. In terms of the constraints, $\complement_1$ to $\complement_4$ are constraints imposed by the hardware limitations of the respective components.  $\complement_5$ to $\complement_8$ ensure that the processing frequency and transmission power of UAV and MDs are always within the effective range. $\complement_9$ and $\complement_{10}$ guarantee that every task is executed at exactly one place at any moment (no duplicate computing). $\complement_{11}$ restricts that the total energy consumption of the UAV during $T$ does not exceed its battery capacity, as $E^{\text{UAV}}(t)$ denotes the energy consumption of the UAV at the $t$th timeslot, 
\begin{equation}
\resizebox{0.9\linewidth}{!}{
    $
    \begin{aligned}
    E^{\text{UAV}}(t) &= E^{\text{MOV}}(t)+\sum_{j=1}^{K} \sum_{i=1}^{N_j} \left[o_{i j}^{\text{DC}}(t) \left(\frac{p_{u}(t) s_{i j}(t) }{C_j(t) r_{(u, D)}(t)}\right)  \right. \\
    &\left. + o_{i j}^{\text{UAV}}(t) \left(\frac{p_{u}(t) s_{i j}(t) }{C_j(t) r_{(u, j)}(t)} + \delta_{u} s_{i j}(t) c_{i j}\left(f(t)\right)^{2}\right)\right].
    \end{aligned}
    $
}
\label{eqn:E_UAV}
\end{equation}
Note that the processing frequencies $f(t)$, $f_{j}(t)$, and transmission powers $p_u(t)$, $p_j(t)$ are dependent on the decision variables $\mathbf{o}_{ij}(t)$.  Numerous scholarly works (e.g.,~\citep{Wei2019}) provided comprehensive analyses of these intricate interactions.

\section{Algorithms}
\label{sec: Algorithm for Trajectory}

The problem~\eqref{eq:problem} possesses an intrinsic nonconvex nature. We divide the problem into three subproblems, one focusing on controlling the attitude, the second on generating an initial trajectory and altitude, and the other on task assignment and resource allocation, as well as further tuning the trajectory, respectively. 

\subsection{Attitude Control} \label{subsec:atc}

\subsubsection{Classical PID Control}
The classical PID control algorithm is a widely used algorithm for UAV attitude control~\citep{Obias2019}. It considers the following three control parts: 

\begin{itemize}
    \item Proportional control (P): the error of the controller input is amplified by a certain proportion. This constitutes the basic control part; 
    \item Integral control (I): 
    an integral term is introduced for correction of the proportional control to maintain the system stability. Its coefficient will approach $0$ as time goes by;
    \item Differential control (D): a differential term based on the rate of change of the input error is added. The purpose is to improve the response accuracy by predicting the future trend of the input error.
\end{itemize}

The classical PID control is described by the following equation, 

\begin{equation}
U(t)=k_{p} e(t)+k_{i} \int e(t) d t+k_{d}\frac{d} e(t){d t},
\end{equation}
where $k_{p}$, $k_{i}$ and $k_{d}$ are the proportional, integral and differential coefficients, respectively. The output of the PID control system, $U(t)$, will affect $\omega, \theta, \phi,\psi$, which will in turns have an impact on the power consumption of the UAV (see Section~\ref{sec:structure} and~\ref{appendixA} for details).  

%%%%%%%%%%%%%%%%%%%%%%%%%%%%%%%%%%%%%%%%%%%%%
%  % *****
\subsubsection{Fuzzy PID Control}
The fuzzy PID control algorithm improves on the classical PID control by adding nonlinear adaptive capability through Fuzzy Set Theory~\citep{Yu2024tvt,Batikan2016}. By inputting fuzzification and a set of fuzzy rules, the PID gain is dynamically adjusted according to the current error $E$ and the error rate $EC$. In addition, the coefficients $k_p$, $k_i$, and $k_d$ are no longer constants, enabling them to adapt to environmental changes. However, since the fuzzy rules are preset based on expert knowledge, the controller lacks flexibility in adapting to unforeseen complex environmental changes in real time~\citep{Shri2024}.

\subsubsection{FEAR-PID Control}
\begin{figure}[htbp]
\centering
\includegraphics[width=\linewidth]{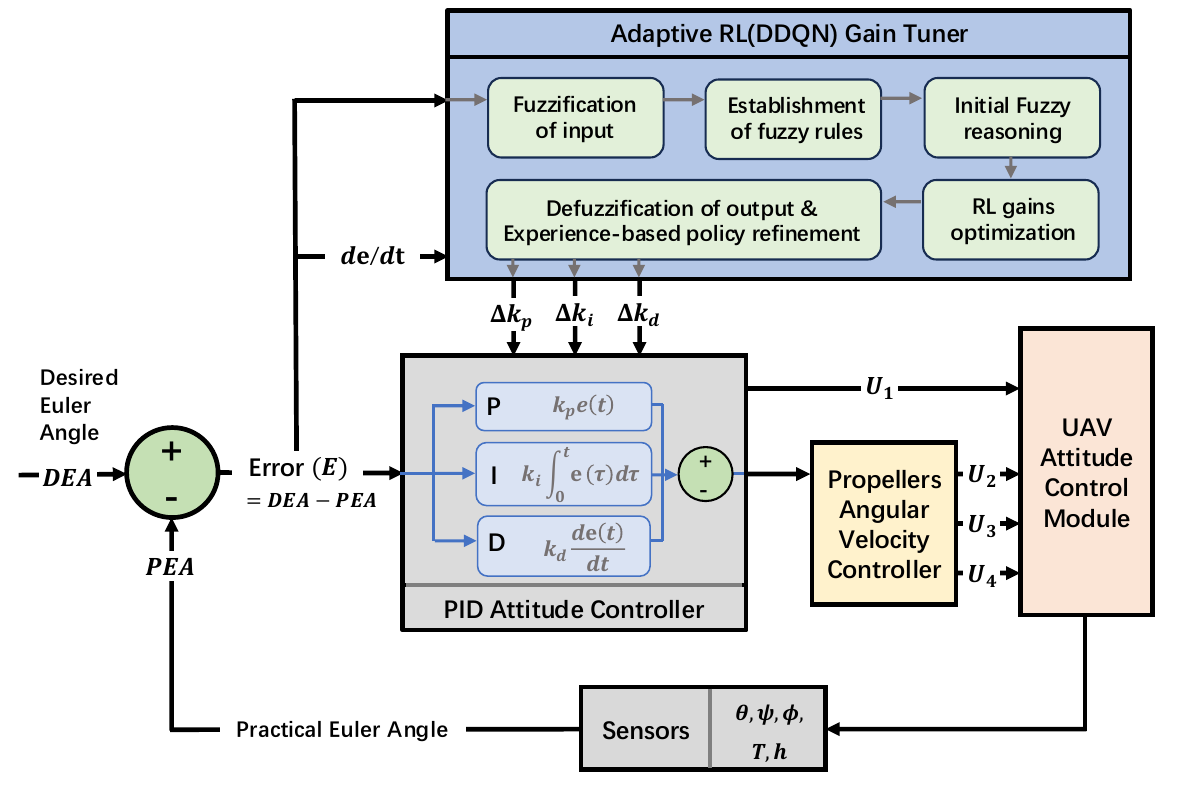}
\caption{Schematic structure of FEAR-PID control system for UAV.}
\label{fig:PID}
%\vspace{-0.5cm} % *****
\end{figure}

To solve the above problems, we propose the Fuzzy-Enhanced Adaptive Reinforcement PID (FEAR-PID) controller.
FEAR-PID integrates RL in fuzzy PID to provide an additional adaptive layer. The control process of FEAR-PID is demonstrated in Fig.~\ref{fig:PID}. Specifically, the initial control strategy is first provided by fuzzy logic based on expert knowledge.
The specific rules are shown in Table~\ref{tab: Fuzzy rules}. 
\begin{table}[!htb]
\begin{center}
\caption{INITIAL FUZZY PID CONTROL RULES} 
\label{tab: Fuzzy rules}
\resizebox{0.99\columnwidth}{!}{
\begin{tabular}{|l|ccccccc|}
% \hline 
\hline 
\multicolumn{8}{|c|}{NB: Negative Big ($-3$)\quad \quad NM: Negative Middle ($-2$) \quad \quad NS: Negative Small ($-1$)\quad \quad ZO (0): zero} \\
\multicolumn{8}{|c|}{PB: positive big ($3$) \quad \quad PM: positive middle ($2$) \quad \quad PS: positive small ($1$)} \\
\Xhline{1.5pt}
& \multicolumn{7}{c|}{\textbf{$EC$}} \\
\cline { 2 - 8 } \textbf{ $E$} & \textbf{NB} & \textbf{NM} & \textbf{NS} & \textbf{ZO} & \textbf{PS} & \textbf{PM} & \textbf{PB} \\
\cdashline { 2 - 8 }[2pt/2pt]
& $\Delta k_{p}$/$\Delta k_{i}$/$\Delta k_{d}$ & $\Delta k_{p}$/$\Delta k_{i}$/$\Delta k_{d}$ & $\Delta k_{p}$/$\Delta k_{i}$/$\Delta k_{d}$ & $\Delta k_{p}$/$\Delta k_{i}$/$\Delta k_{d}$ & $\Delta k_{p}$/$\Delta k_{i}$/$\Delta k_{d}$ & $\Delta k_{p}$/$\Delta k_{i}$/$\Delta k_{d}$ & $\Delta k_{p}$/$\Delta k_{i}$/$\Delta k_{d}$ \\
\hline 
\textbf{NB} & PB/NB/PS & PB/NB/NS & PM/NM/NB & PM/NM/NB & PS/NS/NB & ZO/ZO/NM & ZO/ZO/PS \\
\textbf{NM} & PB/NB/PS & PB/NB/NS & PM/NM/NB & PS/NS/NM & PS/NS/NM & ZO/ZO/NS & NS/PS/ZO \\
\textbf{NS} & PM/NB/ZO & PM/NM/NS & PM/NS/NM & PS/NS/NM & ZO/ZO/NS & NS/PM/NS & NS/PM/ZO \\
\textbf{ZO} & PM/NM/ZO & PM/NM/NS & PS/NS/NS & ZO/ZO/NS & NS/PS/NS & NM/PM/NS & NM/PM/ZO \\
\textbf{PS} & PS/NM/ZO & PS/NS/ZO & ZO/ZO/ZO & NS/PS/ZO & NS/PS/ZO & NM/PM/ZO & NM/PB/ZO \\
\textbf{PM} & PS/ZO/PB & ZO/ZO/PM & NS/PS/PM & NM/PS/PM & NM/PM/PS & NM/PB/PS & NB/PB/PB \\
\textbf{PB} & ZO/ZO/PB & ZO/ZO/PM & NM/PS/PM & NM/PM/PM & NM/PM/PS & NB/PB/PS & NB/PB/PB \\
\hline
% \hline
\end{tabular}}
\end{center}

\end{table}

Then, the PID gain coefficients in the fuzzy controller are obtained by a Double Deep Q-Network (DDQN). The RL module continuously optimizes the PID gain coefficients based on real-time feedback, thus effectively adapting to the dynamic environment and improving the control effect. The RL agent in FEAR-PID is designed as follows:

\begin{itemize}
    \item \textbf{State:} The state $S_t$ at timeslot $t$ consists of the $E$, $EC$, and the PID coefficients $\Delta k_p$, $\Delta k_i$, and $\Delta k_d$. Represented as $S_t = [E, EC, \Delta k_p, \Delta k_i, \Delta k_d]$.
    
    \item \textbf{Action:} The action $A_t = [\Delta k_p, \Delta k_i, \Delta k_d]$ represents the adjustments made to the PID coefficients at each step, resulting in updated PID gains,
    \begin{equation}
    \left\{\begin{array}{l}
    k_{p-FEAR }= k_{p}+\Delta k_{p} \\
    k_{i-FEAR }= k_{i}+\Delta k_{i} \\
    k_{d-FEAR }= k_{d}+\Delta k_{d}.
    \end{array}\right.
    \end{equation}
    
    \item \textbf{Reward:} The reward function $R_t$ is designed to guide the controller toward minimizing the control error and achieving stable responses. 
    \begin{equation}
    R_t = -\left( \vartheta_1 |E| + \vartheta_2 \left| \frac{dE}{dt} \right| + \vartheta_3 \cdot \Delta U^2 + \vartheta_4 \cdot \Upsilon \right)
    \end{equation}
    where the absolute error $|E|$ aims to minimize the error. The term $\left| \frac{dE}{dt} \right|$ controls the rate of error change for stability, $\Delta U^2$ captures control output variation, incentivizing smoother adjustments, AND $\Upsilon$ denotes the overshoot penalizing excessive responses and aids in maintaining stability. The weights $\vartheta_1, \vartheta_2, \vartheta_3,$ and $\vartheta_4$ are provided to balance accuracy, convergence speed, stability, and control smoothness. The specific values and ranges of relevant parameters %required by the Mamdani reasoning method for fuzzy PID control~\citep{Sakti2014} 
    summarized in Table~\ref{tab: FEAR-PID PARAMETER SETTINGS}.
\end{itemize}

\begin{table}[htbp]
\caption{FEAR-PID PARAMETER SETTINGS \label{tab: FEAR-PID PARAMETER SETTINGS}}
\begin{center}
\resizebox{0.99\columnwidth}{!}{
\begin{tabular}{|c;{2pt/2pt}c|c;{2pt/2pt}c|c;{2pt/2pt}c|}
\hline 
Parameters & Values & Parameters & Values & Parameters & Values \\
\hline 
\multirow{2}{*}{\centering $E$, $EC$} & -6, -5, -4, -3, -2, -1, & \multirow{2}{*}{\centering $\Delta k_{p}$, $\Delta k_{d}$ }& -3, -2, -1, 0, & 
\multirow{2}{*}{\centering $\Delta k_{i}$} & -1.5, -1, 0, \\
& 0, 1, 2, 3, 4, 5, 6 &  & 1, 2, 3 &  & 1, 1.5 \\
\hline 
$\vartheta_1$ & 3 & $\vartheta_2$ & 0.5 & $\vartheta_3, \vartheta_4$ & 1.5 \\
\hline
\end{tabular}
}
\end{center}
\end{table}

%%%%%%%%%%%%%%%%%%%%%%%%%%%%%%%%%%%%%%%%%%%%%%%%%%%%%%%%

\subsection{Task Assignment and Resource Allocation}

%We propose two algorithms to address the two subproblems, respectively, with enhanced decoupling and safety values mechanism incorporated in the second algorithm to overcome the deadlock issue in conventional heuristic algorithms. 

\begin{figure*}[t]
\centering
\includegraphics[width=0.75\textwidth]{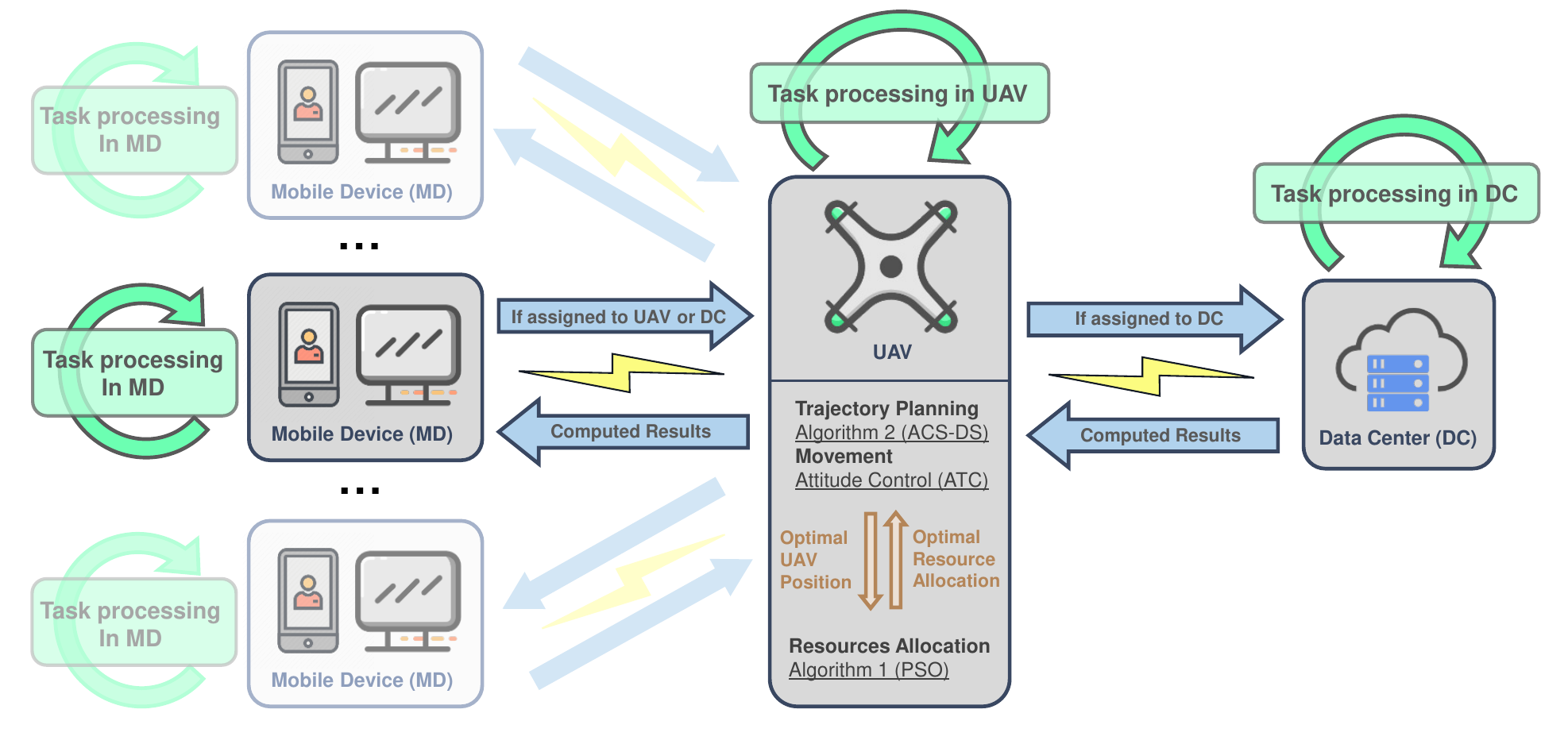}
\caption{A flowchart demonstration of the components in our proposed framework.}
\label{fig:Flowchart}
\end{figure*}

An overview of the steps in our proposed framework to address the task assignment, resource allocation and trajectory planning modules in the joint optimization problem is shown in Fig.~\ref{fig:Flowchart}. We will apply an Ant Colony System based algorithm (Algorithm~\ref{Algorithm2}, to be described in detail later) to determine an initial trajectory based on the terrain of the area and the position of MDs. Then, the attitude control mechanism described in Section~\ref{subsec:atc} will be invoked to determine the as well as tune the trajectory. Finally, a Particle Swarm Optimization based algorithm (Algorithm~\ref{Algorithm1}, which was initially proposed in the conference version~\citep{Liu2022}) will determine the optimal task assignments and resource allocations. We will also analyze the convergence performance of the proposed algorithms at the end of the section.

We first introduce an algorithm based on the idea of Particle Swarm Optimization (PSO), for assigning tasks and allocating resources given that the trajectory is already known. The steps of our proposed algorithm are demonstrated in Algorithm~\ref{Algorithm1}, which we refer the readers to the conference version of this paper~\citep{Liu2022} for more detailed explanations. We categorize the job assignment decision, transmission power, and processing frequency as a collective entity  $s_h = (\mathbf{o}_{ij}, p_j(t), f_j(t))$ within the particle swarm. Here, $h \in \{1, ..., H\}$. Next, the determination of the particle group's position is carried out by uniformly sampling a set of particle groups denoted by $H\in\mathbb{N_+}$. In order to search for the optimal solution, the velocity of the particle groups is initialized and subsequently updated using the method in~\citep{Khan2018}. The algorithm terminates when the discrepancy between the outcomes of $20$ successive rounds is less than a specified threshold $\eth$, and then outputs the minimal value $\mathbb{S}^{*}$, the optimal decision for job assignment $\mathbf{o}_{ij}^{*}$, the transmission power $p_j^*(t)$, and the processing frequency $f_j^*(t)$.

\revise{Regarding fairness considerations, we note that fairness is not explicitly included as an optimization target in our objective function, which focuses on minimizing overall energy consumption and delay. However, our system design incorporates natural fairness mechanisms through orthogonal channel allocation (preventing channel monopolization) and the feasibility constraints that ensure all tasks are assigned to appropriate execution locations. The Gamma-based channel allocation strategy further promotes equitable resource distribution by preventing extremely large tasks from dominating available channels. A comprehensive quantitative fairness analysis, including metrics such as Jain's fairness index or per-user delay variance, would require additional modeling extensions and is considered as future work for scenarios where explicit fairness guarantees are required.}

\renewcommand{\algorithmicrequire}{\textbf{Input:}}  
\renewcommand{\algorithmicensure}{\textbf{Output:}}
\begin{algorithm}[ht] 
	\caption{Task Assignment, Power and Frequency Allocations based on Particle Swarm Optimization} 
    \begin{algorithmic}[1] 
	\Require The length of an interval $t$, %$p_j(t)$, $f_j(t)$, 
  $p^{\mathrm{max}}_j(t)$; $p^{\mathrm{min}}_j(t)$; $f^{\mathrm{max}}_j(t)$; $f^{\mathrm{min}}_j(t)$; $F_{A1}$; $F_{A2}$; $F_{I}$ %\timothy{changed, check if correct}
		\Ensure $\mathbb{S}^{*}(t)$; $\mathbf{o}_{ij}^{*}$; $p_j^{*}(t)$; $f_j^{*}(t)$
		%%\vspace{1 ex}
		\For {$h = 1$ to $H$}
			\State Initialize particles' positions  $G_h$ %\timothy{deleted $x,y,z$ to avoid confusion}
			\State Initialize particles' velocities $V_h$ 
			\State Initialize $pBest_h \longleftarrow G_h$
		\EndFor
	    \State Initialize $gBest(0)$ $\longleftarrow$ argmin$fit(pBest_h)$, where $fit$ represent the equation  to compute $\mathbb{S}^{*}(t)$; $k = 1$
	    \For{$h=1$ to $H$}
			\State Update $V_h$ and $G_h$ by acceleration factors  $F_{A1}$, $F_{A2}$ and $F_{I}$
			\If{$fit(G_h) < fit(pBest_h)$}
 			   \State $pBest_h \longleftarrow G_h$
	            \If{$fit(pBest_h)<fit(gBest)$}
 		        \State $gBest(k) \longleftarrow pBest_h$
 		   	    \EndIf
	        \EndIf
	        \State $k \longleftarrow (k+1)$
		\EndFor
	   % \EndWhile
	   \While{$\lvert gBest(k+1)-gBest(k) \rvert < \eth$}
	    \State $\mathbb{S}^{*}(t) \longleftarrow fit(gBest)$
	    \State $(\mathbf{o}_{ij}^{*}$; $p_j^{*}(t)$; $f_j^{*}(t))$ $\longleftarrow gBest$
        \EndWhile
	    \State Output $\mathbb{S}^{*}(t)$; $(\mathbf{o}_{ij}^{*}$; $p_j^{*}(t)$; $f_j^{*}(t))$
        
	\end{algorithmic} \label{Algorithm1}
	%%\vspace{-0.1cm}
\end{algorithm}

%%%%%%%%%%%%%%%%%%%%%%%%%%%%%%%%%%%%%%%%%%%%%%
%%\vspace{-0.1cm} % *****
\subsection{ACS-DS for Trajectory Planning: Motivation and Concept}
%%\vspace{-0.1cm} % *****

We now present the concept and outlines of the ACS-DS (Ant Colony System with \textbf{Decoupling} and \textbf{Safety values}) algorithm to generate the optimal UAV trajectory to minimize $E^{\mathrm{MOV}}$. Then, we will prove that ACS-DS always converges to the optimal solution in polynomial time in subsequent subsections.

%%\vspace{-0.2cm} % *****

%Our objective at this stage is to design a computationally efficient algorithm, based on the Ant Colony System Algorithm (ACS), that can generate the optimal UAV trajectory to minimize $E_{u}$. 

We first divide the 3D space into discrete grids, with the center of each unit of the grid serving as a \textit{waypoint}, that is, the position that the UAV will pass. The 3D terrain is generated randomly, and we denote the area at and below terrain surfaces, referred as the \textit{no-fly zone}, by $\boldsymbol O = \{(x^o_n, y^o_n, z^o_n)\}$, where $n \in \{1, 2, ..., N_{o}\}$, and $N_{o}$ is the total number of grids in the no-fly zone. Note that the classical ACS is not applicable to our three-dimensional trajectory planning problem as it struggles to converge in path searching in large three-dimensional spaces, and is likely to be trapped in local optima. 

To overcome these issues and obtain the optimal trajectory with minimum cost in an efficient manner, we propose the ACS-DS, by incorporate the two special mechanisms to the classical ACS.  We now elaborate these two mechanisms. 

\subsubsection{Safety Values}
%\timothy{better change $\mu$, $\nu$, $k$ here as they were used for other meanings before, how about use $\mu$, $\nu$, and $\iota$ instead?} 
We enhance the heuristic function of the ACS-DS by adding a security value at each step based on the number of currently known the numbers of feasible and infeasible waypoints. Specifically, we determine the safety value of a waypoint based on the proportion of known feasible waypoints in the next available position for that waypoint. The safety value from waypoint $\mu$ to $\nu$ is calculated as $\kappa_{\mu\nu}=\frac{N_{f}-N_{\mu\nu}}{N_{f}}$, where $N_f$ denotes the total number of waypoints in a preset constant range $\left[0, R_{f}\right]$, and $N_{\mu\nu}$ denotes the number of infeasible waypoints in the range of $N_f$ over direction from $\mu$ to $\nu$. Safety values are updated using a similar rule to pheromones, and when choosing its next move, the ant would add the safety values to the pheromone levels for all possible actions. We will show in the results section that this method can reduce the running time and improve the operation efficiency compare to the classical ACS algorithm.

%%\vspace{0.1cm} % *****
\subsubsection{Decoupling}
In ACS, when an ant reaches a point with no further viable options, it becomes trapped in a deadlock. To address this issue, we introduce a mechanism that allows the ant to escape the deadlock. Specifically, when any of following rule is satisfied during the ant's movement, the Decoupling mechanism will be triggered to perform a backtracking behavior with depth (step size) $D_b$ to find alternative directions:
\begin{itemize}
    \item An ant repeats a closed cycle consisting of two or more waypoints over multiple consecutive timeslots.
    \item A sudden drop in the amount of pheromone and safety values occurs, as indicated by the fact that the $\kappa$ of the currently selected waypoint is less than half that of the previous waypoint. 
    \item An ant falls into a local optimum and undetected in the early stage, that is, the ant has not triggered backtracking behavior for more than 25 consecutive waypoints in the first one-third of all iterations.
\end{itemize}

It is worth noting that the third rule is only considered in the early stages of the algorithm (first one-third of all iterations) to avoid excessive backtracking that reduces the convergence speed of the algorithm. This mechanism, by dynamically adjusting the pheromone levels and allowing the ant to backtrack and explore new possibilities, ensures that the ACS-DS remains robust and capable of finding optimal trajectories even in complex and challenging scenarios.

% \vspace{-0.8cm} % *****
\subsection{ACS-DS: Detailed Steps}
We let $m$ be the total number of ants in the colony, and $\sigma=V_P+\kappa$ be the guidance factor. We further denote $\sigma_{\mu\nu}(h)$ as the sum of pheromone values and safety values between the neighboring waypoints $\mu$ and $\nu$ in the $h$th iteration. %, with a maximum value of $g(s^*)$. 
The initial pheromones on each edge are equal, namely $\sigma_{\mu\nu}(0)=C$ for all $\mu$ and $\nu$. We denote by $s^*$ the global best path obtained by the algorithm, corresponding to the minimal-cost trajectory in our UAV optimization problem. Next, for each ant $k \in \{1,2\ldots,m\}$ in the colony, we initialize the pheromone $V_{P0}$, the safety value $\kappa_{0}$, and the heuristic value $\eta$. We also define the evaporation rate $\rho\in(0,1)$ representing the degree to which the guidance factor $\sigma_{\mu\nu}(h)$ decays with iterations. %\timothy{do you need the subscript $\iota$ for these variables?} \timothy{two $\rho$ ?} \revise{There's only one $\rho$. I made a mistake here. The two sentences should go together.} \timothy{need a subscript representing the current iteration?}. 
Finally, we define $\mathscr{S}_{\iota}$ as the set of waypoints that ant $\iota$ can pass next, and $p_{\mu\nu}^{\iota}(h)$ as the probability that ant $\iota$ moves from position $\mu$ to $\nu$ in the $h$th iteration. The exact value of $p_{\mu\nu}^{\iota}(h)$ is jointly determined by the guidance factor and heuristic value at the waypoint, as in the following equation,
\begin{equation}\label{eqn: ACS_probability}
p_{\mu\nu}^{\iota}(h)= \begin{cases}\frac{\sigma_{\mu\nu}^\alpha(h) \eta_{\mu\nu}{ }^\beta(h)}{\sum_{r\in \mathscr{S}_{\iota}} \sigma_{\mu r}^\alpha(h) \eta_{\mu r}{ }^\beta(h)} & \nu \in \mathscr{S}_{\iota}, \\ \quad \quad \quad 0 & \text { otherwise.}\end{cases}
\end{equation}

Detailed steps of ACS-DS are presented in Algorithm~\ref{Algorithm2}.

\renewcommand{\algorithmicrequire}{\textbf{Input:}} 
\renewcommand{\algorithmicensure}{\textbf{Output:}} 
\begin{algorithm}[ht]
%\vspace{-0.05cm} % *****
        \small
	\caption{ACS-DS: ACS-based Trajectory Planning with Decoupling and Safety Values Mechanisms}
	\begin{algorithmic}[1]
		\Require Positions of all MDs $(x_j,y_j,z_j)$;$\rho$; $\eta$; $V_{P0}$; $\kappa_{0}$; terrain surface with no-fly zone $(x^o_n, y^o_n, z^o_n) \in \boldsymbol O$ %\timothy{moved from first statement to input}
		\Ensure $E^{\mathrm{MOV}*}(t)$; Trajectory of the UAV $(x_\iota(t),y_\iota(t),z_\iota(t))$ %\timothy{changed notation for consistency with the previous section}
		%\vspace{1 ex}
		%\State Randomly generate terrain surfaces (no-fly zone) in the map, record the position as coordinates $(x^o_n, y^o_n, z^o_n) \in \boldsymbol O$
        \For{$h = 1$ to $H$}
            \State Randomly set the initial coordinate of the UAV, $(x(0),y(0),z(0))$
          \While{$(x(0),y(0),z(0)) \notin \boldsymbol O$}
		\For{each edge}
			\State Set initial pheromone, and calculate the  initial safety values
		\EndFor
		\For{each ant $\iota$}
			\State $(x_\iota(h),y_\iota(h),z_\iota(h))$ = $(x(0),y(0),z(0))$
	    \For{each edge $(\mu,\nu)$}
	            \State Choose the next coordinate with probability $p_{\mu\nu}^{\iota}(h)$ by $\sigma$ and $\eta$
	        \While{$(x_\iota(h+1),y_\iota(h+1),z_\iota(h+1)) \notin \boldsymbol O$}
	    	\State Output $(x_\iota(h+1),y_\iota(h+1),z_\iota(h+1))$
                \EndWhile
		\EndFor
		\State Compute and output the length $\sum_{t=1}^{T} d_{u}(t)$ of  the path by the $\iota$th ant and $E_{u}$ 
		\For {each edge $(\mu,\nu)$}
	       \State Update $V_{P_{\mu\nu}}$ and $\kappa_{\mu\nu}$ by $\rho$
              \State Update $\sigma_{\mu\nu}(h)$ by $V_{P_{\mu\nu}}$ and $\kappa_{\mu\nu}$
            \EndFor
            \EndFor
          \EndWhile
        \EndFor
        \State Compute and output $E^{\mathrm{MOV}*}(t)$ by \eqref{eqn: move_enegy}.
        %\State \lsjnew{Call Algorithm \ref{Algorithm1} to compute and output $\mathbb{E}^*(t)$ by \ref{eqn:total_energy}.}
	\end{algorithmic}  \label{Algorithm2}
\end{algorithm}

%\vspace{-0.2cm} % *****
\subsection{ACS-DS: Proof of convergence} 
We now prove that Algorithm~\ref{Algorithm2} will eventually converge, starting by the following proposition. 

\begin{proposition} \label{pro: acs1}
In Algorithm~\ref{Algorithm2}, for the guidance factor $\sigma_{\mu\nu}$ on any edge $(\mu,\nu)$ generated by the ants during the searching process, there exists a maximum value $g(s^*)$ as $h \to \infty$.
\end{proposition}

\begin{proof}
For ACS-DS, the local update of the guidance factor for edge $(\mu,\nu)$ after the completion of each round of iteration can be represented as $\sigma_{\mu \nu}(h+1)=(1-\varphi) \cdot \sigma_{\mu \nu}(h)+\varphi \cdot \Delta \sigma_{\mu \nu}(h)$, while the global guidance factor is updated in a similar way, that is, $\sigma(h+1)=(1-\varphi) \cdot \sigma(h)+\varphi \cdot \Delta \sigma(h)$. Here, $\Delta \sigma_{\mu \nu}(h) = \sigma_{\mu \nu}(h) - \sigma_{\mu \nu}(h-1)$ is the increment of guidance factor on edge $(\mu,\nu)$ at the $h$th iteration.

%and the global update of the guidance factor after the completion of each round of iteration can be uniformly represented as, $\sigma_{\mu \nu}(h+1)=(1-\varphi) \cdot \sigma_{\mu \nu}(h)+\varphi \cdot \Delta \sigma_{\mu \nu}(h),$ where $\varphi$ stands for heuristic value $\eta$ or evaporation rate $\rho$, and $\Delta \sigma_{\mu \nu}(h)$ is the guidance factor increment on each edge after each iteration, has a maximum value of $g(s^*)$. Therefore, the maximum values of guidance factor for each generation are,

\begin{equation}
\begin{aligned}
& \sigma_{\mu \nu}(1)=(1-\varphi) \cdot \sigma_{\mu \nu}(0)+\varphi \cdot \Delta \sigma_{\mu \nu}(0) \\
%& \sigma_{\mu \nu}(2)=(1-\varphi) \cdot \sigma_{\mu \nu}(1)+\varphi \cdot \Delta \sigma_{\mu \nu}(1) \\
%& \hspace{0.87cm}=(1-\varphi)\left[(1-\varphi) \cdot \sigma_{\mu \nu}(0)+\varphi \cdot \Delta \sigma_{\mu \nu}(0)\right]+\varphi \cdot \Delta \sigma_{\mu \nu}(1)\\
%& \hspace{0.87cm}=(1-\varphi)^2 \cdot \sigma_{\mu \nu}(0) + \varphi \cdot(1-\varphi) \cdot \Delta \sigma_{\mu \nu}(0)+\varphi \cdot \Delta \sigma_{\mu \nu}(1) \\
& \cdots \\
& \sigma_{\mu \nu}(h)=(1-\varphi)^h \cdot \sigma_{\mu \nu}(0) + (1-\varphi)^{h-1} \cdot \varphi \cdot \Delta \sigma_{\mu \nu}(1) + \cdots\\
& \quad\quad\quad\quad +(1-\varphi) \cdot \varphi \cdot \Delta \sigma_{\mu \nu}(h-1) +\varphi \cdot \Delta \sigma_{\mu \nu}(h) \\
& \hspace{0.87cm} =\sum_{q=1}^h\left((1-\varphi)^{h-q} \cdot \varphi \cdot \Delta \sigma_{\mu \nu}(q) + (1-\varphi)^h \cdot \sigma_{\mu \nu}(0)\right)\\
& \lim _{h \rightarrow \infty} \sigma_{\mu \nu}^{\max }(h)=\lim _{h \rightarrow \infty}\left(\sum_{q=1}^h\left((1-\varphi)^{h-q} \cdot \varphi \cdot \Delta \sigma_{\mu \nu}(q) + \cdots \right) \right)\\
&\quad \quad \quad \quad \quad=g(s^*) < \infty.
\end{aligned}
\end{equation}

Thus, the guidance factor on each edge is bounded from the above by $g(s^*)$.
\end{proof}

After the first optimal solution is found, the guidance factor on the elements belonging to an optimal solution is guaranteed to be no less than that on other elements as we have a sufficient number of subsequent generations. That is, the guidance factor on any element not belonging to an optimal solution will keep decreasing until it is no larger than the guidance factor on the elements belonging to an optimal solution. In mathematical terms, we have the following corollary.

\begin{corollary} \label{cor: acs1}
As $h \to \infty$, it is always true that $\sigma_{\mu \nu}(h) \geq \sigma_{\mu'\nu'}(h)$, if $(\mu, \nu) \in s^*$ and $(\mu', \nu') \notin s^*$.
\end{corollary}

\begin{proof} Suppose it takes $h^*$ generations to find the first optimal solution. 
We assume that for a certain $(\mu,\nu) \in s^*$, there exists a $(\mu',\nu') \notin s^*$, $\sigma_{\mu\nu}\left(h^*\right)<\sigma_{\mu'\nu'}\left(h^*\right)$. Then, based on the guidance factor update rule in ACS-DS, by the $h^*+h^{\prime}$ generation, the guidance factor on $(\mu',\nu')$ can be derived as
\begin{equation}
\sigma_{\mu'\nu'}\left(h^*+h^{\prime}\right)=\max \left\{\sigma_{\mu\nu}\left(h^*\right),(1-\rho)^{h^{\prime}} \cdot \sigma_{\mu'\nu'}\left(h^*\right)\right\}
%\end{aligned}
\end{equation}

Therefore,
\begin{equation}
\begin{aligned}
\lim _{h \rightarrow \infty} \sigma_{\mu'\nu'}\left(h^*+h^{\prime}\right)
&=\max \Bigl\{ \sigma_{\mu\nu}\left(h^*\right), \\
&\qquad \lim _{h \rightarrow \infty}\left[(1-\rho)^{h^{\prime}} \cdot \sigma_{\mu'\nu'}\left(h^*\right)\right]\Bigr\} \\
&=\max \left\{\sigma_{\mu\nu}\left(h^*\right), 0\right\} \\
&\le \sigma_{\mu\nu}\left(h^*\right).
\end{aligned}
\end{equation}
\end{proof} 
Proposition 1 and Corollary 1 jointly prove that, after a sufficient number of iterations, the guidance factor on an optimal path is bounded and no less than those on other paths. Noticeably, the backtracking behavior caused by the decoupling mechanism in ACS-DS will only increase the number of iterations for the current waypoint, but does not affect the updating of the guidance factor and the convergence to the optimal solution as in the original ACS. Therefore, the ACS-DS algorithm always converges to the global optimal solution.

\subsection{ACS-DS: Complexity analysis}
We now analyze the complexity of Algorithm~\ref{Algorithm2} when a total of $n$ users initiate computation requests in the current timeslot. Assume that in the worst case, Algorithm~\ref{Algorithm2} triggers backtracking at every waypoint for each ant upfront, the path construction time complexity for $m$ ants is $O\left(m \cdot n^2 \right)$. The complexity of each update of the guidance factor, and each calculation of the transfer probability is $O\left(n^2\right)$. If the algorithm is finished after $H$ iterations, the complexity of the entire algorithm is $O\left(H \cdot m \cdot n^2 + H \cdot n^2\right)$. Since the magnitude of $m$ is similar to $n$, the complexity of the ACS-DS algorithm in completing all the tasks of the network is approximately $O\left(H \cdot n^3\right)$, which means that Algorithm~\ref{Algorithm2} has a polynomial time complexity. 

%\vspace{-0.1cm} % *****
\section{Performance Evaluation} \label{sec: experiment}
%\vspace{-0.1cm} % *****
We now present numerical results to assess the efficacy and versatility of our proposed solutions.
\subsection{Experiment setup} \label{subsec: experiment_setup}
\begin{figure*}[t]
    \centering
    \begin{subfigure}{0.45\textwidth} % 让图片共享一行
        \centering
        \includegraphics[width=\linewidth]{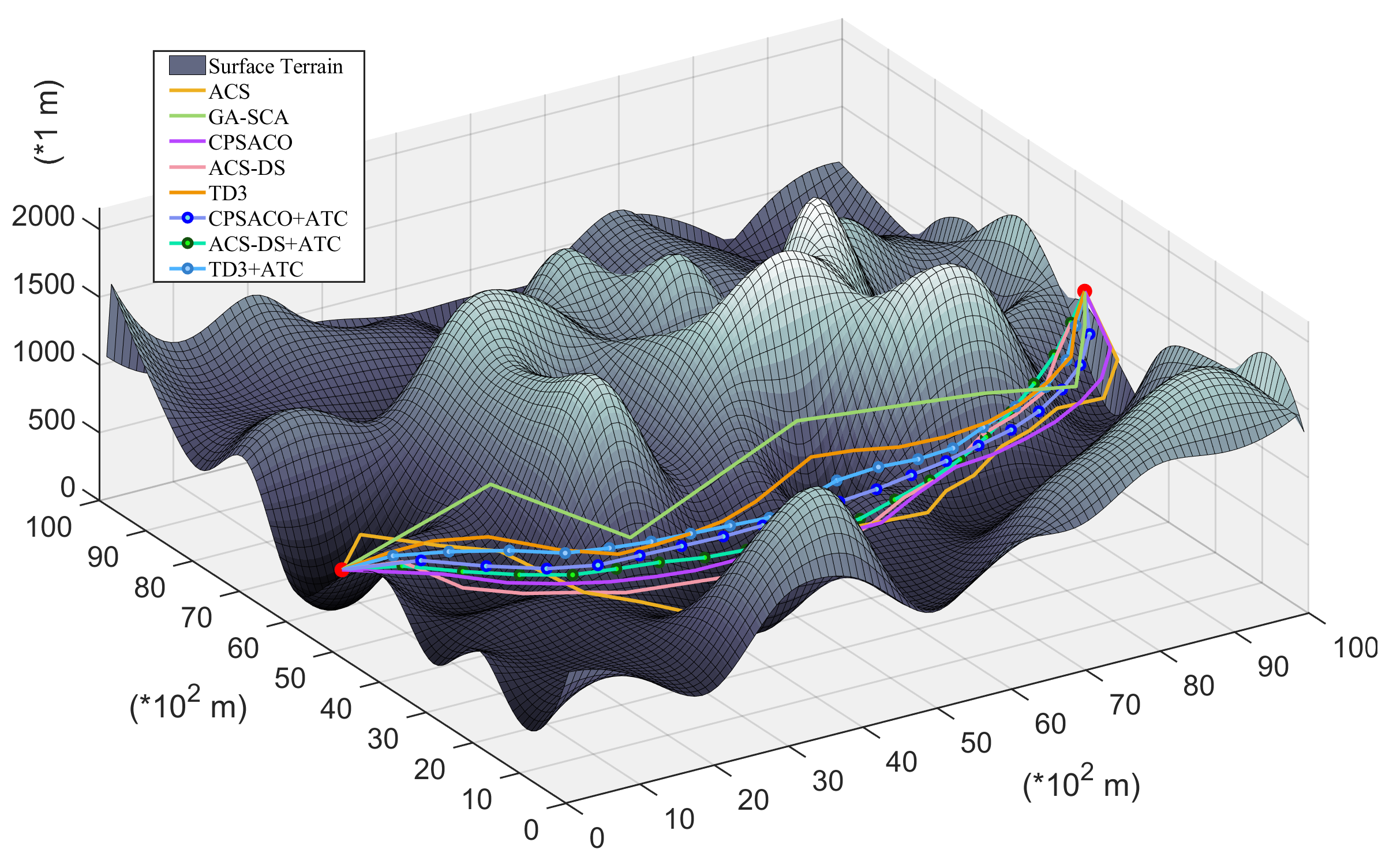}
        \caption{Side view}
        \label{fig:four path}
    \end{subfigure}
    \hfill
    \begin{subfigure}{0.45\textwidth} % 让第二张图片靠右对齐
        \centering
        \includegraphics[width=\linewidth]{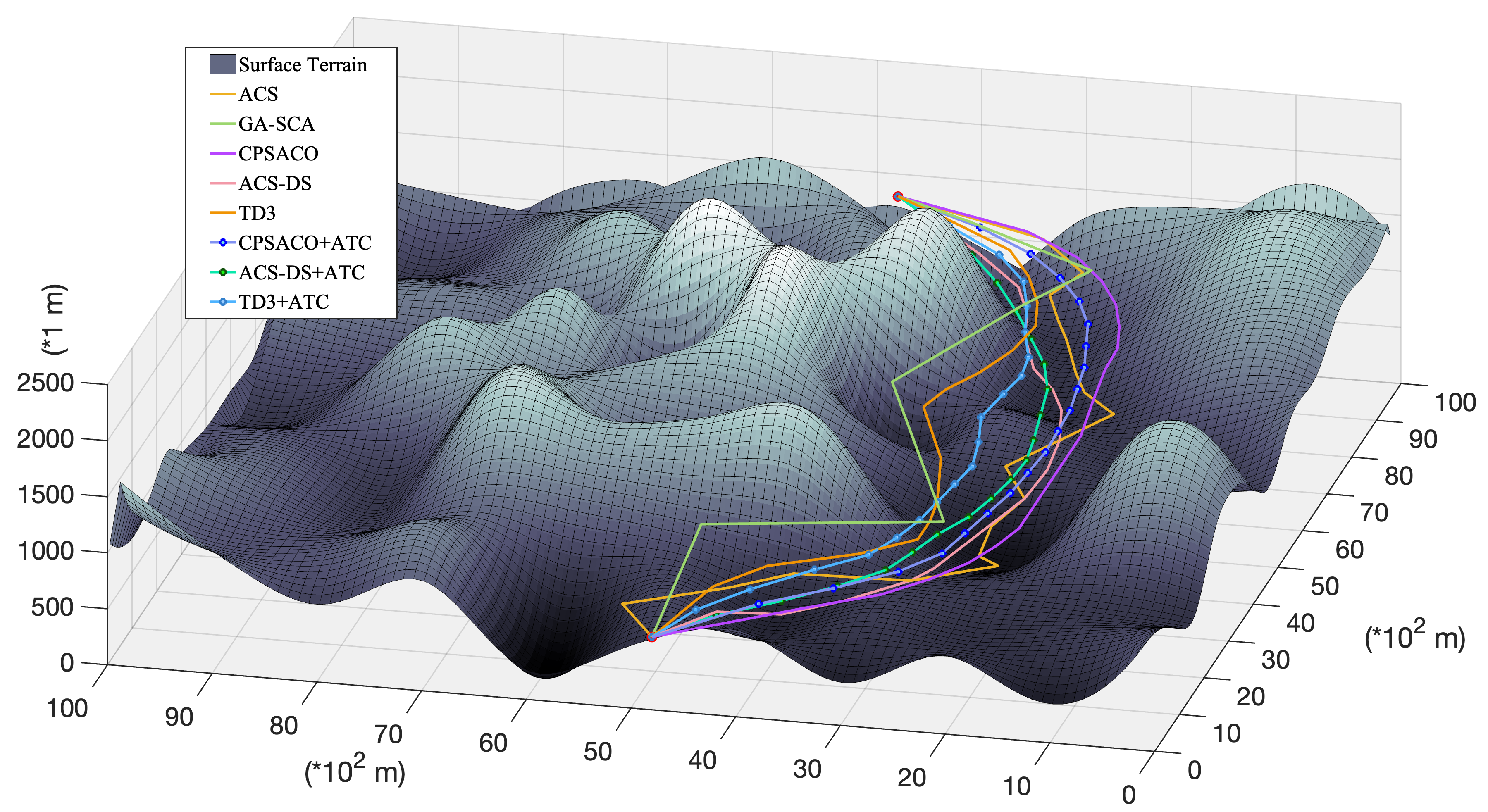}
        \caption{Top view}
        \label{fig:four path1}
    \end{subfigure}
    \caption{Trajectories planned by different mechanisms on a 3D spatial domain. \revise{The terminal point shown in Fig. 7 represents the UAV's final hovering location for sustained communication rather than the physical position of the MD.}}
    \label{fig:four_paths_combined}
\end{figure*}
We consider a three-dimensional spatial domain with dimensions of $S \times S \times Z$. We randomly generate a continuous surface (no-fly zone) within a three-dimensional spatial domain as shown in Figs.~\ref{fig:four path} and~\ref{fig:four path1}, and consider that $K$ MDs are placed on the ground conforming to the terrain height, with their positions uniformly and randomly distributed in the space. \revise{All MDs are placed at the terrain surface height, and the UAV always operates at an altitude higher than the MD elevations.} The values of key system parameters are listed in Table~\ref{tab:MODEL PARAMETER SETTINGS}. 
% \vspace{0.5cm} % *****
\begin{table}[htbp]
\caption{VALUES OF KEY SYSTEM PARAMETERS
\label{tab:MODEL PARAMETER SETTINGS}}
\begin{center}
\resizebox{0.98\columnwidth}{!}{
\begin{tabular}{|c;{2pt/2pt}c|c;{2pt/2pt}c|}
\hline 
\textbf{Parameter} & \textbf{Value} & \textbf{Parameter} & \textbf{Value}
\\
\hline 
$S$ & 50000 $\mathrm{~m}$ & $Z$ & 3000  $\mathrm{~m}$\\
\hline
$L$  & 0.1$\mathrm{~s}$  & $\epsilon$ & [0.05, 1.00] \\
\hline
$m$ & \revise{50 $\mathrm{~kg}$} & $a$ & 5.28 $\mathrm{~m} / \mathrm{s}^2$ \\
\hline
$g$ & 9.81 $\mathrm{~m}/ \mathrm{s}^2$ & $l$ & 1150 $\mathrm{~mm}$ \\
\hline
$K$ & 50 & $R$ & 512 $\mathrm{~mm}$ \\
\hline
$N_{c}$ & 40 & $P_t$ & 7.5 $\mathrm{~mm}$ \\
\hline
$N_{B}$ & 2 & $P_{w}$ & 250 $\mathrm{~mm}$ \\
\hline
$B_c$ & 5000 $\mathrm{~Wh}$ & $I_x$ & \revise{5.5} \\
\hline
$v^{\max}$ & \revise{15$\mathrm{~m/s}$} & $I_y$ & \revise{5.5} \\
\hline
$z^{\max}$ & 2500 $\mathrm{~m}$ & $I_z$ & \revise{10.0} \\
\hline
\revise{$\omega^{\min}$} & \revise{200 rpm} 
& \revise{$\omega^{\max}$} & \revise{5000 rpm} \\
\hline
$C_{T}$ & 1.483 $~\left(\mathrm{N} \cdot \mathrm{m} \cdot \mathrm{min}^2\right) / \mathrm{r}^2$ 
& $C_{M}$ & 2.925 $~\left(\mathrm{N} \cdot \mathrm{m} \cdot \mathrm{min}^2\right) / \mathrm{r}^2$ \\
\hline
$M_{g}$ & 29 $\mathrm{~g/mol}$
& $R_{g}$ & 8.314 $\mathrm{~J/(mol*K)}$ \\
\hline
\revise{$T$} & \revise{20–35 min} 
& \revise{$N_j$} & \revise{1–5} \\
\hline
\revise{$f^{\min}$} & \revise{1.0 GHz} & \revise{$f^{\max}$} & \revise{3.0 GHz} \\
\hline
\revise{$f_j^{\min}$} & \revise{0.5 GHz} & \revise{$f_j^{\max}$} & \revise{2.0 GHz} \\
\hline
\revise{$\delta_u$} & \revise{$1.2\times10^{-28}$ J/(cycle$^2\cdot$bit)} & \revise{$\delta_j$} & \revise{$3.0\times10^{-28}$ J/(cycle$^2\cdot$bit)} \\
\hline
\revise{$\bar{Q}$} & \revise{$5\times10^{7}$ (bit-equivalent)} & \revise{$\tau$} & \revise{1.0 s} \\
\hline
\end{tabular}
}
\end{center}
\end{table}
\revise{The UAV flight time $T$ is obtained by cumulatively summing the propulsion energy consumed at each timeslot, where the energy usage is determined from the rotor-speed–dependent motor model already defined in Section~\ref{sec:structure}. The endurance is reached once the accumulated consumption equals the battery capacity $B_c$, giving the maximum feasible $T$ under a given trajectory and control strategy. Specifically, $T$ is calculated as:
\begin{equation}
T = \max \left\{ t \in \mathbb{N} : \sum_{\tau=0}^{t} E^{\text{UAV}}(\tau) \leq B_c \right\},
\label{eqn:T_calculation}
\end{equation}
where $E^{\text{UAV}}(\tau)$ is defined in~\eqref{eqn:E_UAV} and includes both movement energy $E^{\mathrm{MOV}}(\tau)$ defined in~\eqref{eqn: move_enegy}) and other energy consumption components.}
Values of key parameters used in the algorithms described earlier in this paper are listed in Table~\ref{tab:ALGORITHM PARAMETER SETTINGS}. All experimental results presented in this section are based on the average of the 50 independent runs with the weighting factor $\epsilon$ randomly generated within its domain for each run.
% \vspace{0.4cm} % *****
\begin{table}[htbp]
\caption{VALUES OF KEY ALGORITHMIC PARAMETERS \label{tab:ALGORITHM PARAMETER SETTINGS}}
\begin{center}
\resizebox{0.98\columnwidth}{!}{
\begin{tabular}{|c;{2pt/2pt}c|c;{2pt/2pt}c|c;{2pt/2pt}c|}
% \hline
\hline \textbf{Parameter} & \textbf{Value} & \textbf{Parameter} & \textbf{Value} & \textbf{Parameter} & \textbf{Value}
\\
\hline
$\rho$ & 0.25
& $V_{P0}$ & 3.8 & $\eta$ & 2.5\\
$F_{A1}$, $F_{A2}$ & 2.0
& $F_{I}$ & 0.65 & $R_f, D_b$ & 200 $~\mathrm{m}$\\
\hline
% \hline
\end{tabular}
% \vspace{-0.2cm} % *****
}
\end{center}
% \vspace{-0.2cm} % *****
\end{table}
For the task assignment and resource allocation part, we adopt PSO (Algorithm~\ref{Algorithm1}), which has been demonstrated effective and robust in our earlier conference paper~\citep{Liu2022}, for all experiments in this section. For the attitude control and trajectory planning modules, we focus on evaluating the performance of our proposed ACS-DS (Algorithm~\ref{Algorithm2}) and ATC (Section~\ref{subsec:atc}). \revise{Note that our work addresses a joint UAV control--communication--computation optimization problem, which fundamentally differs from traditional networking-only scheduling tasks. Conventional networking baselines such as Round Robin, Proportional Fair, or queue-based schedulers cannot operate in our setting, as they assume fixed infrastructure and do not interact with UAV flight dynamics, propulsion energy, or real-time orientation-dependent channel variations. These classical methods also do not model UAV physical constraints, motor-level energy consumption, computation frequency control, or end-to-end latency decomposition, all of which are essential components of our optimization problem. Therefore, consistent with the standard practice in the UAV-assisted MEC and cross-layer optimization literature, we adopt representative algorithmic families that can jointly optimize continuous control variables and discrete resource allocation decisions.} Specifically, we examine the following ten implementations, including our proposed approach, recently proposed heuristic (e.g.,~\citep{Zheng2024GASCA,Yan2024CPSACO}), and RL-based methods~\citep{Tan2024TD3}, and compare their performances in terms of overall operational efficiency cost.

\begin{itemize}
\item \textbf{ACS}: The trajectory of the UAV is planned using the classical Ant Colony System (ACS). Task assignment, processing frequencies, transmission powers, and channel allocation follow the optimization procedures described in Section~\ref{subsec: Network}. No special mechanisms are applied for attitude control.

\item \textbf{GA-SCA}: The UAV trajectory is optimized using a hybrid method combining Genetic Algorithm (GA) and Successive Convex Approximation (SCA), as proposed in \citep{Zheng2024GASCA}. The remaining parameters follow the same strategy as ACS.

\item \textbf{CPS-ACO}: The UAV trajectory is planned using the Chaotic-Polarized-Simulated Ant Colony Optimization (CPS-ACO) method, proposed in \citep{Yan2024CPSACO}. Other system parameters are optimized as in ACS.

\item \textbf{TD3}: The UAV trajectory is planned using the Twin Delayed Deep Deterministic Policy Gradient (TD3) algorithm, as described in \citep{Tan2024TD3,Zheng2024TD3}. Other configurations are consistent with the baseline setup in ACS.

\item \textbf{ACS-DS}: This is an enhanced version of ACS, where UAV trajectory planning incorporates our proposed decoupling and safety value mechanisms, detailed in Algorithm~\ref{Algorithm2}. Other parameter settings remain unchanged.

\item \textbf{CPS-ACO+ATC}: Based on CPS-ACO, this method integrates the attitude control mechanism described in Section~\ref{subsec:atc} to enable more stable UAV dynamics.

\item \textbf{TD3+ATC}: Based on TD3, this variant incorporates the attitude control mechanism described in Section~\ref{subsec:atc} to improve stability and responsiveness.

\item \textbf{ACS-DS+ATC}: This is the full version of our proposed method, which combines the decoupled safe trajectory planning in ACS-DS with the attitude control mechanism in Section~\ref{subsec:atc}.
\end{itemize}

\subsection{Simulation Environments}
\revise{\begin{sloppypar}
The experimental evaluation of the proposed framework is conducted using three complementary simulation environments, each designed to address different aspects of system validation and serving distinct roles in the evaluation pipeline. 

\begin{figure*}[t]
    \centering
    \begin{subfigure}{0.492\linewidth}
        \centering
        \includegraphics[width=\linewidth]{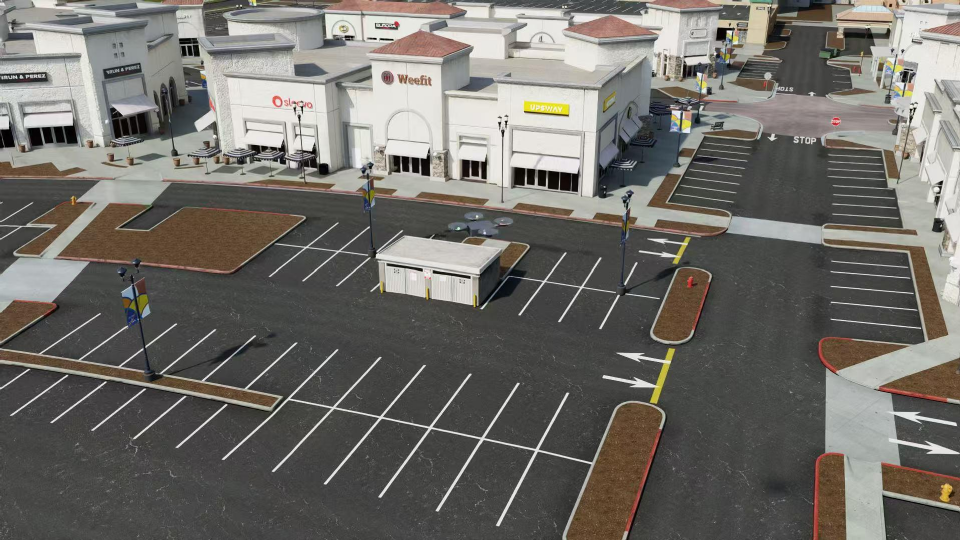}
        \caption{Isaac sim deployment simulation}
        \label{fig:isaac}
    \end{subfigure}
    \hfill
    \begin{subfigure}{0.485\linewidth}
        \centering
        \includegraphics[width=\linewidth]{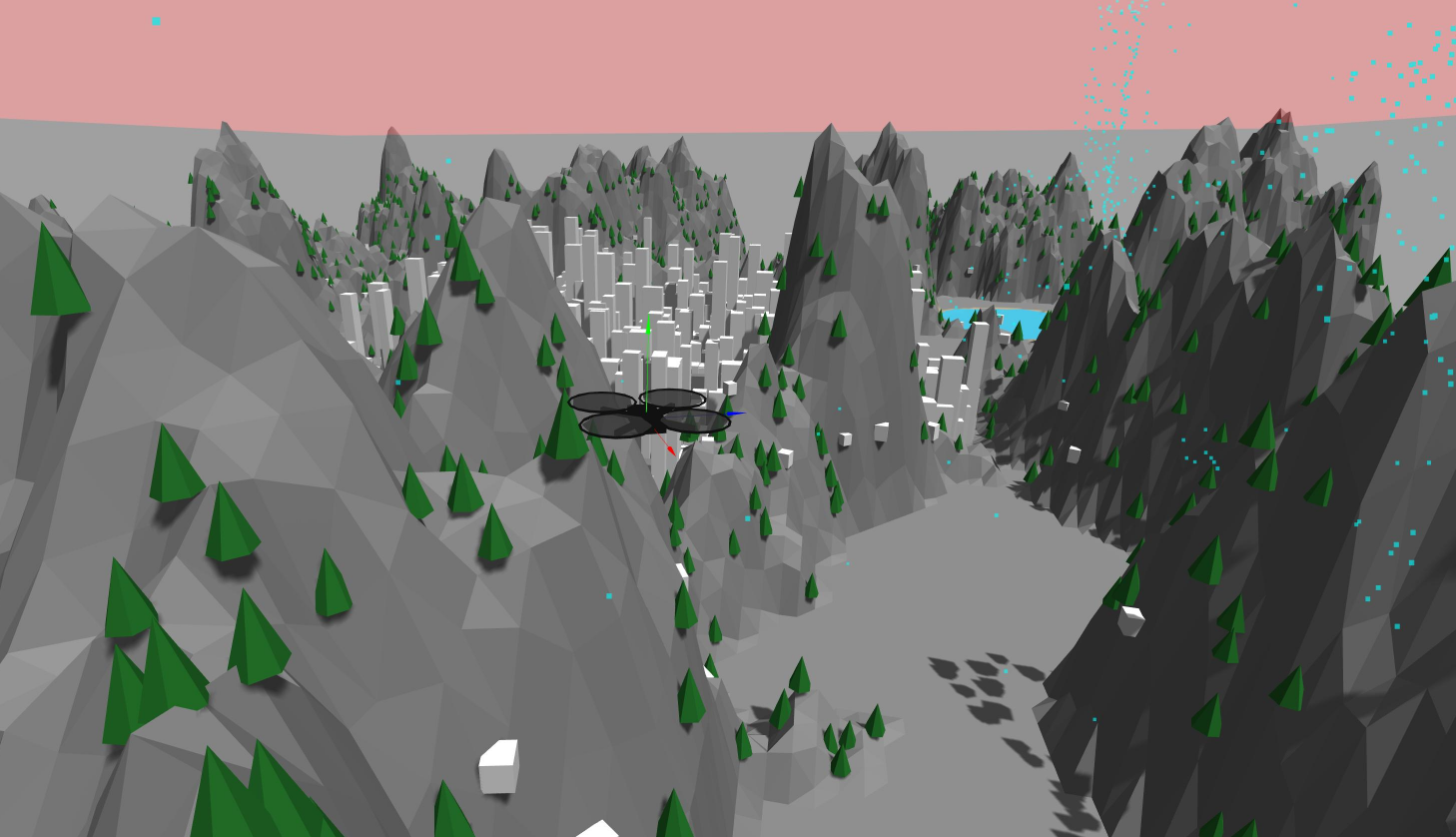}
        \caption{Web-based demonstration}
        \label{fig:web_demo}
    \end{subfigure}
    \vspace{0.5cm}
    \begin{subfigure}{0.49\linewidth}
        \centering
        \includegraphics[width=\linewidth]{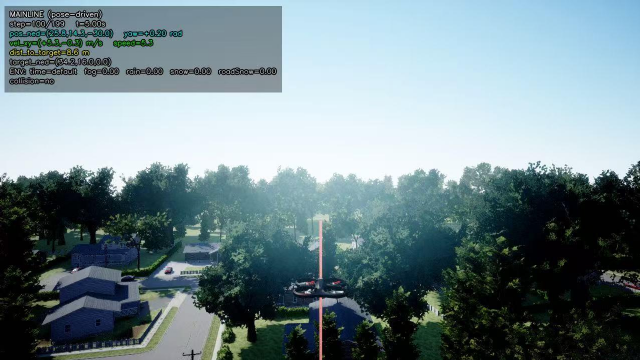}
        \caption{AirSim deployment simulation (AirSimNH)}
        \label{fig:airsim1}
    \end{subfigure}
    \hfill
    \begin{subfigure}{0.492\linewidth}
        \centering
        \includegraphics[width=\linewidth]{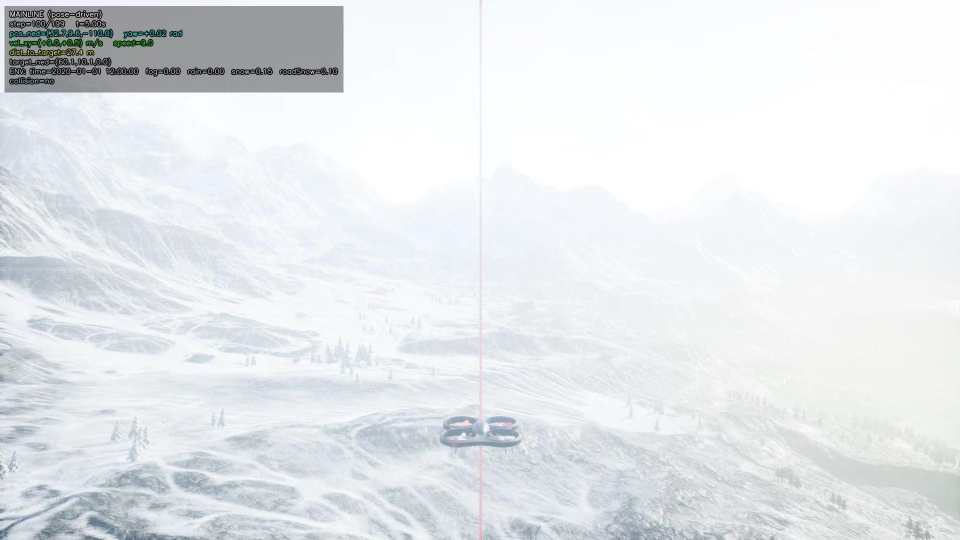}
        \caption{AirSim deployment simulation (Landscape Mountains with snow)}
        \label{fig:airsim2}
    \end{subfigure}
    \caption{Simulation environments used for framework evaluation.}
    \label{fig:simulation_environments}
\end{figure*}

\begin{itemize}
    \item \textbf{Isaac Sim}: This Omniverse/PhysX-based robotics simulation platform provides high-fidelity rigid-body UAV dynamics that accurately model inertia, thrust, and torque characteristics. As shown in Fig.~\ref{fig:isaac}, the environment incorporates realistic environmental disturbances including wind fields, actuator latency, and platform vibration, enabling high-fidelity physical verification of control algorithms. We utilize Isaac Sim to validate the FEAR-PID controller's stability under realistic dynamics and to verify that the UAV parameters specified in Table~\ref{tab:MODEL PARAMETER SETTINGS} exhibit physically correct behavior. We additionally leverage Isaac Sim's built-in/official USD environment assets as realistic scene data for representative outdoor scenarios, so that the controller is also inspected under non-trivial geometry and terrain context beyond synthetic setups. This physics-based validation ensures that our control strategies and parameter selections are feasible for real-world deployment.
    
    \item \textbf{Microsoft AirSim}: This platform offers photorealistic rendering powered by the Unreal Engine, featuring realistic suburban and mountainous environments that represent the large-scale outdoor deployment scenarios described in our system model (e.g., AirSimNH and LandscapeMountains scenes). As shown in Fig.~\ref{fig:airsim1} and Fig.~\ref{fig:airsim2}, the environment provides high-fidelity visual representation of complex terrain with realistic lighting and atmospheric effects. The platform provides comprehensive sensor emulation including RGB cameras, depth sensors, semantic segmentation, IMU, GPS, and realistic noise models, and also exposes environment controls such as time-of-day and weather. We employ AirSim to support interactive mission execution with the built-in multirotor dynamics under deployment-scale scenes and configurable conditions, and to generate synchronized FPV/chase-view visualizations for representative runs via ExternalPhysicsEngine (pose-driven) replay, which facilitates consistent presentation and debugging across different methods.
    
    \item \textbf{Lightweight web-based demonstration}\footnote{Online demo: \textcolor{blue}{\url{https://shuaijun-liu.github.io/UAV-Assisted-Fog-Computing-Simulation-Demo}}}: This accessible interactive visualization tool is inspired by AirSim's visual style. As illustrated in Fig.~\ref{fig:web_demo}, the simplified environment employs streamlined UAV dynamics without full rigid-body physics, yet incorporates dynamic wind effects, simplified control loops, and procedural terrain generation. The web demo serves to qualitatively visualize UAV behavior, trajectory evolution, and controller stability, offering practitioners and researchers an intuitive understanding of the system's operational characteristics without requiring specialized simulation software.
\end{itemize}

These three environments complement each other: Isaac Sim validates physical feasibility and control robustness with high-fidelity dynamics and asset-based scenes, AirSim evaluates deployment-scale performance in interactive Unreal Engine environments, and the web demonstration provides intuitive qualitative insights. Together, they form a comprehensive evaluation framework that spans from low-level control validation to high-level mission assessment.
\end{sloppypar}}

\subsection{Module-wise Ablation Study}
To evaluate the individual contribution of each proposed component in our holistic framework, we conduct an ablation study by selectively removing one module at a time while keeping the others unchanged. This experimental design allows us to quantify the performance degradation caused by the absence of specific functionalities and to demonstrate the necessity of integrating all components for optimal system performance. We consider the following four key modules:
\begin{itemize}
    \item \textbf{FEAR-PID-based Attitude Control (ATC)}: Enhances UAV stability and reduces energy overhead during flight maneuvers;
    \item \textbf{ACS-DS Trajectory Planner}: Provides obstacle-aware and energy-efficient path planning with fast convergence. Determining the operational efficiency costs of the movement;
    \item \textbf{PSO-based Task and Resource Assignment}: Allocates task execution and communication resources effectively. Determining the operational efficiency costs of tasks computation and unloading.;
    \item \textbf{Gamma-based Channel Allocation Strategy}: Prevents bandwidth monopolization by larger tasks. Determining the operational efficiency costs of transmission.
\end{itemize}
Each of these modules is removed individually to create a controlled ablation variant of the full framework. The tested ablation variants are summarized in Table~\ref{tab:ablation_variants}.

\begin{table}[htbp]
\centering
\caption{Ablation variants with specific module removals.}
\label{tab:ablation_variants}
\resizebox{0.98\columnwidth}{!}{
\begin{tabular}{|l|l|}
\hline
\textbf{Variant Name} & \textbf{Description} \\
\hline
\textbf{Full} & All modules enabled (ATC + ACS-DS + PSO + Gamma) \\
\textbf{w/o ATC} & Replaces FEAR-PID with classical PID control \\
\textbf{w/o ACS-DS} & Replaces ACS-DS with TD3 trajectory planner \\
\textbf{w/o PSO} & Uses random task assignment and uniform power allocation \\
\textbf{w/o Gamma} & Uses uniform channel assignment ($P_j(t)=1/K$) \\
\textbf{Baseline} & Classical PID + TD3 trajectory + \\ & uniform task/power/channel assignment \\
\hline
\end{tabular}
}
\end{table}

For each variant, we use the same network environment and task configuration as in the full model experiments. All experiments are repeated 50 times with randomized task generation to ensure statistical robustness. The evaluation metric is the \textbf{overall operational efficiency cost}, defined as the sum of total task delay and energy consumption, normalized across scenarios.

%%%%%%%%%%%%%%%%%%%%%%%%%%%%%%%%%%%%%%%%%%%%%%%%
%\vspace{-0.1cm} % *****
\subsection{Numerical results}

\subsubsection{Optimal Propeller Parameters}
\begin{figure}[htbp]
    %\vspace{-0.2cm} % *****
    \centering
    \includegraphics[width=\linewidth]{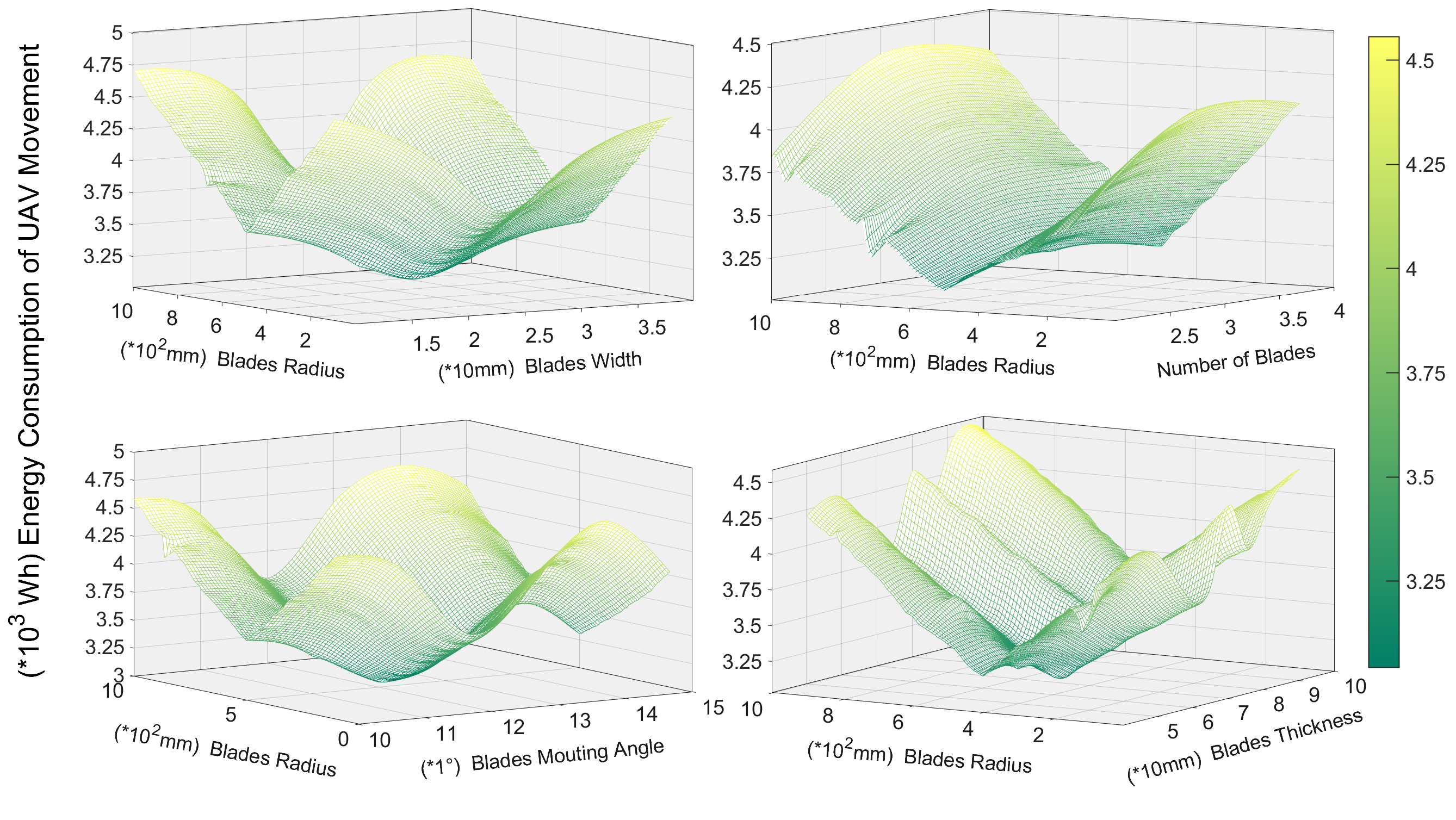}
    \caption{Overall energy consumption vs. propeller blades parameters.}
    \label{fig: Adjustment}
    %\vspace{-0.2cm} % *****
\end{figure}

Fig.~\ref{fig: Adjustment} shows the relationship between the overall energy consumption and propeller blade parameters including the number of blades, and the radius, width, mounting angle, and thickness of each blade, given that all other parameters are fixed. From the results, we can infer that, for a symmetric quadrotor UAV with a total mass of $80~\mathrm{kg}$, four rotors and a maximum acceleration up to $5.28 ~\mathrm{m} / \mathrm{s}^2$, it is optimal to equip $2$ propeller blades with a radius of $500~\mathrm{mm}$, a width of $250~\mathrm{mm}$ and a thickness of $7.5~\mathrm{mm}$. The average energy consumption under configurations with optimal values of parameters can be reduced by more than $30\%$, compared to the average amount of 50 sets of random parameters generated uniformly within the respective allowable range of each parameter.

\vspace{0.4cm}

\subsubsection{Convergence Performance} 
\begin{figure}[htbp]
    %\vspace{-0.2cm} % *****
    \centering
    \includegraphics[width=\linewidth]{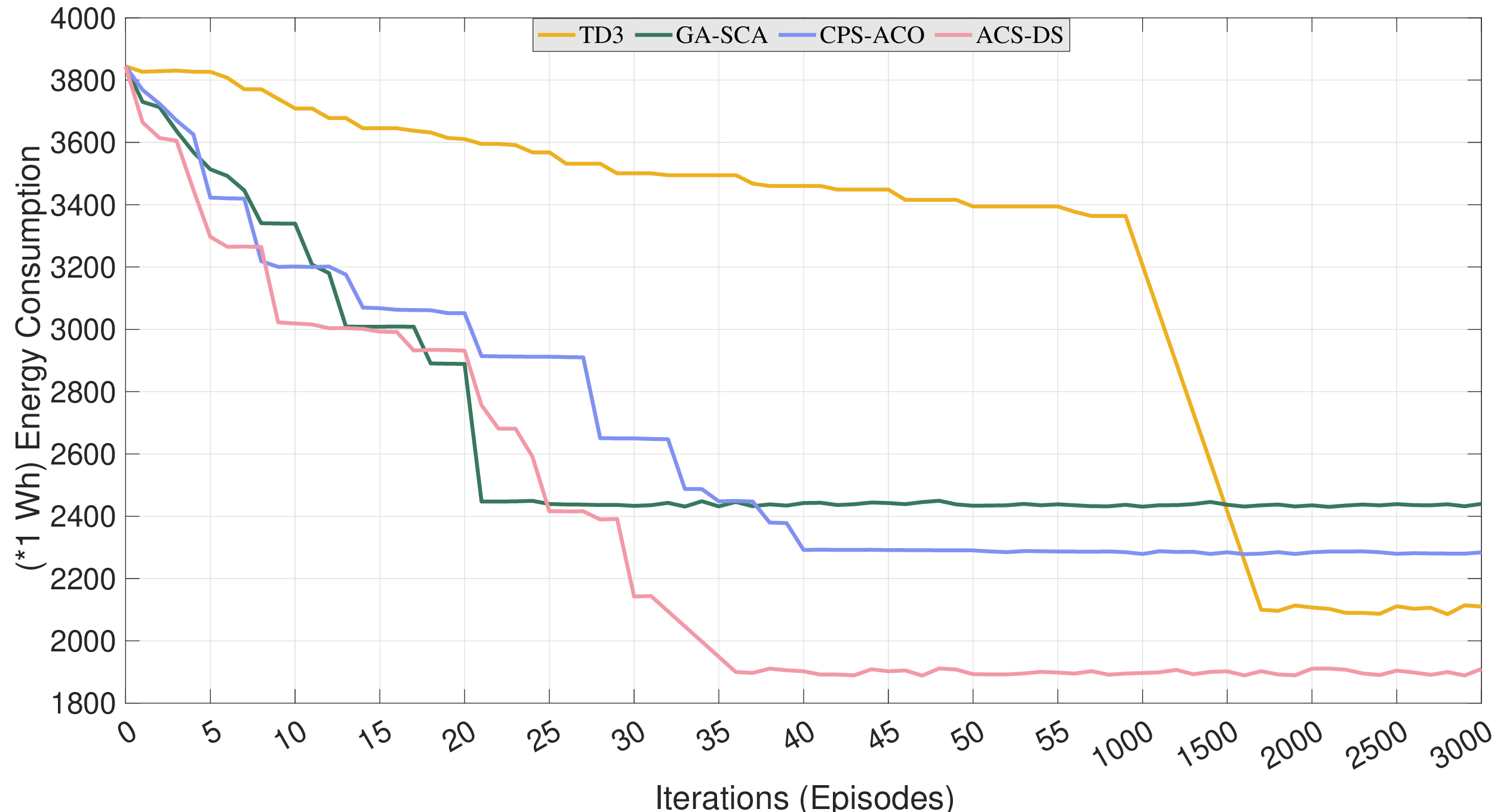}
    \caption{Convergence performances of trajectory planning algorithms.}
    \label{fig: Convergence}
    %\vspace{-0.2cm} % *****
\end{figure}

Fig.~\ref{fig: Convergence} demonstrates the average convergence speeds of GA-SCA,  CPS-ACO, and ACS-DS over $60$ iterations, and TD3 algorithm over $3000$ episodes for $200$ independent runs. The results show that both ACS-DS and TD3 can optimize the energy efficiency through efficient exploring optimal trajectories. However, the average training time per episode for the RL-based TD3 are significantly higher than the average running time per iteration of the other three heuristic algorithms. Therefore, due to its faster convergence and lower computational overhead, ACS-DS is more suitable for rapid decision-making and trajectory planning in unknown environments.

\vspace{0.4cm}

% After the optimal propeller parameters are determined as in the previous subsection, we simulate  the actual rotational speeds and corresponding energy consumption of four rotors across a total of $10$ timeslots \timothy{is 10 too small?}, with the results shown in Fig.~\ref{fig: speedmotors}. It demonstrates that the quadrotor UAV achieves climbing and steering by attitude control within ten randomly intercepted timeslots, which ultimately achieves the purpose of avoiding obstacles and adjusting the transmission position. 

% \begin{figure}[htbp]
%     %\vspace{-0.2cm} % *****
%     \centering
%     \includegraphics[width=8.8cm]{speedmotors.eps}
%     \caption{Motor speed control}
%     \label{fig: speedmotors}
%     %\vspace{-0.2cm} % *****
% \end{figure}

\subsubsection{Trajectory planning with attitude control}
%We then consider the effect of obstacles and test the performance of our trajectory planning algorithm.
%For the 3D region, we randomly generate obstacle height data, and then make the image edges smoother and more consistent with the actual quadrotor UAV application scenario through 2D quadruple convolution interpolation.
From the trajectories shown in Figs.~\ref{fig:four path} and \ref{fig:four path1}, we observe that, the CPS-ACO, TD3, and ACS-DS algorithms are more capable of avoiding no-fly zones. In addition, our proposed Attitude Control (ATC) method can improve the trajectories planned by all algorithms, by effectively correcting unwanted directional changes. Specifically, the FEAR-PID controller dynamically optimizes flight attitude by adjusting the UAV rotor speed in real-time based on sensor feedback, including error and environmental data, resulting in more stable and precise flight control.
%We now generate the terrain height randomly as a continuous surface (no-fly zone) within the three-dimensional spatial domain as shown in Figs.~\ref{fig: four path} and~\ref{fig: four path1}, and consider that the MDs are placed directly on the ground conforming to the terrain height, in order to test the obstacle avoidance capability of our proposed trajectory planning and attitude control algorithms. We observe from the trajectories of the UAV in Figs.~\ref{fig: four path} and \ref{fig: four path1} that, both the additional mechanisms of decoupling and safety values in ACS and attitude control (ATC) have positive impacts on the final path. Specifically, ATC achieves more stable and precise flight control by optimizing the UAV rotor speed in real time based on the feedback of error and environmental information from the sensors. By jointly applying the ACS with the decoupling, safety values and attitude control, ACS-DS+ATC obtains the smoothest and shortest trajectory among all. 

\subsubsection{Operational Efficiency Cost}
\begin{figure}[htbp]
    %\vspace{-0.2cm} % *****
    \centering
    \includegraphics[width=\linewidth]{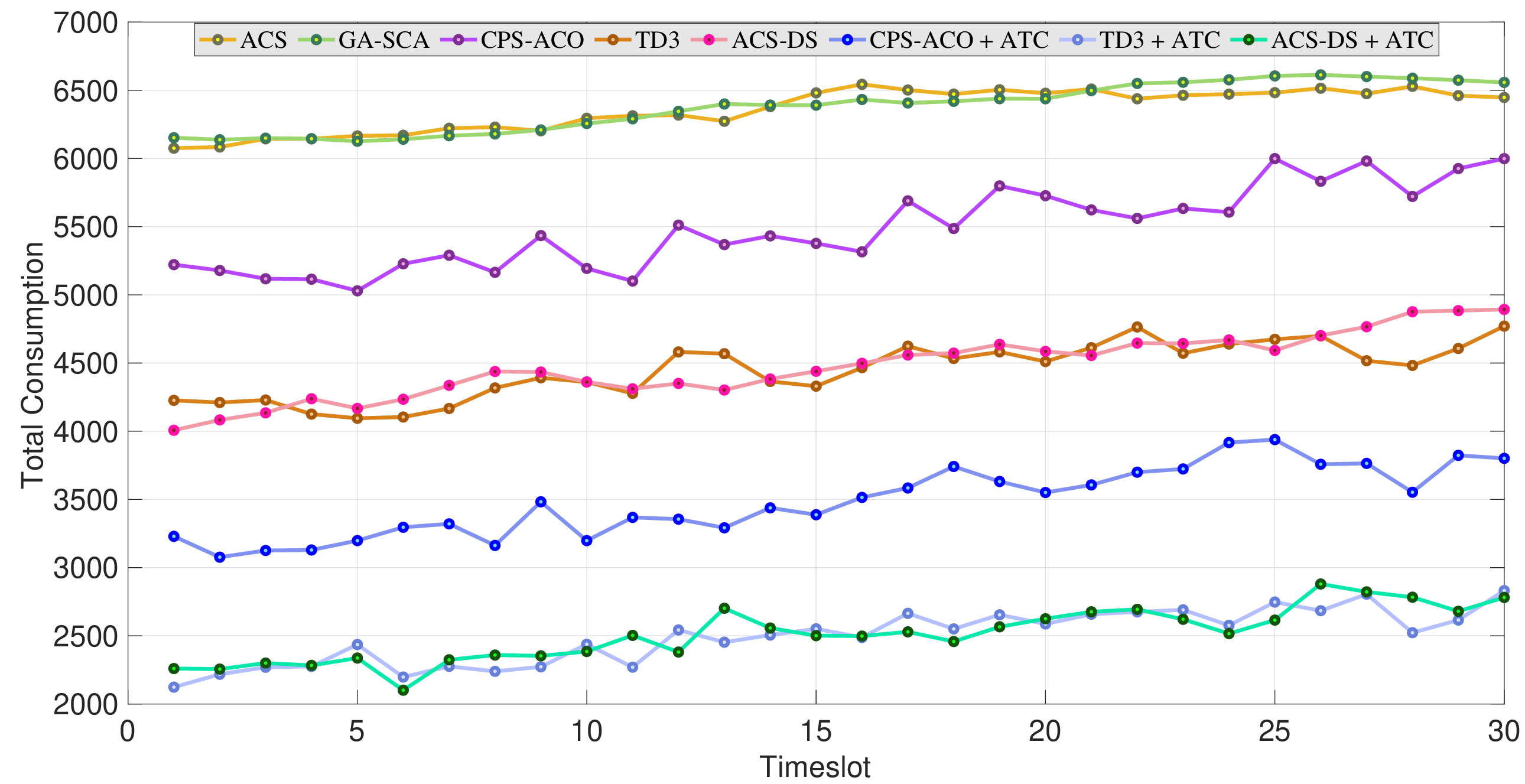}
    \caption{Comparison of operational efficiency cost.}
    \label{fig: result1}
    %\vspace{-0.2cm} % *****
\end{figure}

We present the results of the operational efficiency cost (considering both delay and energy consumption) achieved by different implementations in Fig.~\ref{fig: result1}. \revise{Note that the horizontal axis in this figure (and similar plots throughout this section) represents discrete control time slots, each of duration $L=0.1$ s as defined in Section~\ref{subsec: Network}, rather than the UAV's actual physical flight time. The overall mission duration $T$ spans 20 to 35 minutes in our experiments, calculated as described in Section~\ref{subsec: experiment_setup} below Table~\ref{tab:MODEL PARAMETER SETTINGS}), and the figures display representative segments of the trajectory for clarity of visualization.} One important observation is that, the decoupling mechanism and safety values in ACS-DS can significantly reduce the overall consumption compared to the classical ACS, and have a slight advantage over SOTA (state-of-the-art) reinforcement learning methods such as TD3.

Moreover, the ATC mechanism further enhances the performance of both RL-based (TD3) and heuristic-based (ACS-DS and CPS-ACO) methods. Overall, ACS-DS+ATC achieves the best performance in terms of operational efficiency, closely followed by TD3+ATC. Specifically, compared with conventional ACS, TD3+ATC reduces the operational efficiency cost by 43.5\%, while ACS-DS+ATC reduces it by 48.1\%.

These results validate our initial claim in this paper, that jointly optimizing attitude control, trajectory planning, and task assignment in UAV-assisted fog computing systems effectively captures the interdependencies among these modules. Such integrated approaches (e.g., TD3+ATC and our proposed ACS-DS+ATC) lead to significant improvements in system performance compared to optimizing each aspect individually.

\subsubsection{Trajectory Stability and Robustness Evaluation}
\revise{To evaluate the stability and robustness of different control and planning algorithms in complex 3D environments, we conducted $50$ independent runs for each method under identical initial conditions and environmental settings. The trajectories were generated by our Framework, validated in Isaac Sim and AirSim, and subsequently rendered in the WebGL visualization environment to provide a clearer and lightweight presentation. In all figures, each polyline corresponds to one complete trajectory obtained from a single run.

Fig.~\ref{fig:trajectory_compare_pid} compares the tracking performance of the classical PID controller (red) and the proposed FEAR-PID controller (blue) along the same reference path. The trajectories generated by PID exhibit noticeable drift and oscillation, especially in densely cluttered urban regions. In contrast, FEAR-PID produces a significantly more compact trajectory cluster with reduced lateral deviation, demonstrating superior attitude stability and higher run-to-run consistency.

Fig.~\ref{fig:trajectory_compare_acs} presents the trajectory distributions of three planning algorithms: ACS (yellow), TD3 (green), and ACS-DS (blue). ACS exhibits the largest dispersion, indicating higher sensitivity to environmental variations. TD3 shows moderate improvement in stability, while ACS-DS produces the most concentrated trajectories among the three, confirming that its decoupling mechanism and dynamic safety factors substantially enhance planning robustness across repeated trials.

Overall, these results show that,
\begin{enumerate}
\item FEAR-PID effectively reduces disturbance-induced drift and improves attitude control precision;  
\item ACS-DS generates more stable and reliable flight paths under repeated execution;  
\item Jointly optimizing control and planning is essential for ensuring dependable UAV operation in realistic 3D environments.
\end{enumerate}

\begin{figure}[htbp]
    \centering
    \includegraphics[width=\linewidth]{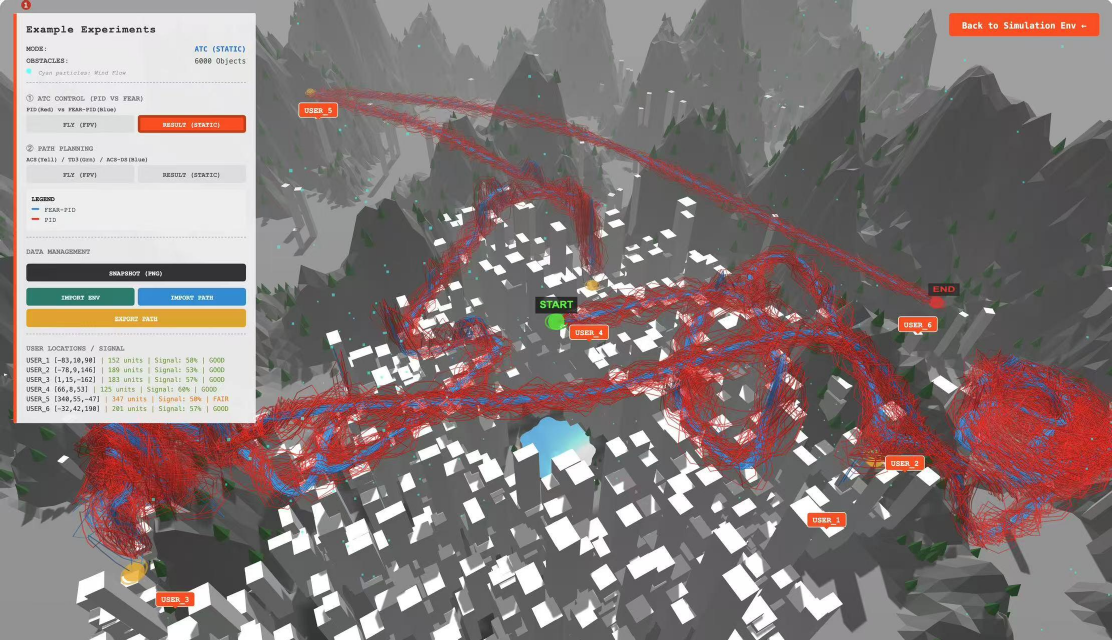}
    \caption{Trajectory comparison between PID (red) and FEAR-PID (blue). Visualization of $50$ independent runs for evaluating trajectory stability and robustness.}
    \label{fig:trajectory_compare_pid}
\end{figure}

\begin{figure}[ht]
    \centering
    \includegraphics[width=\linewidth]{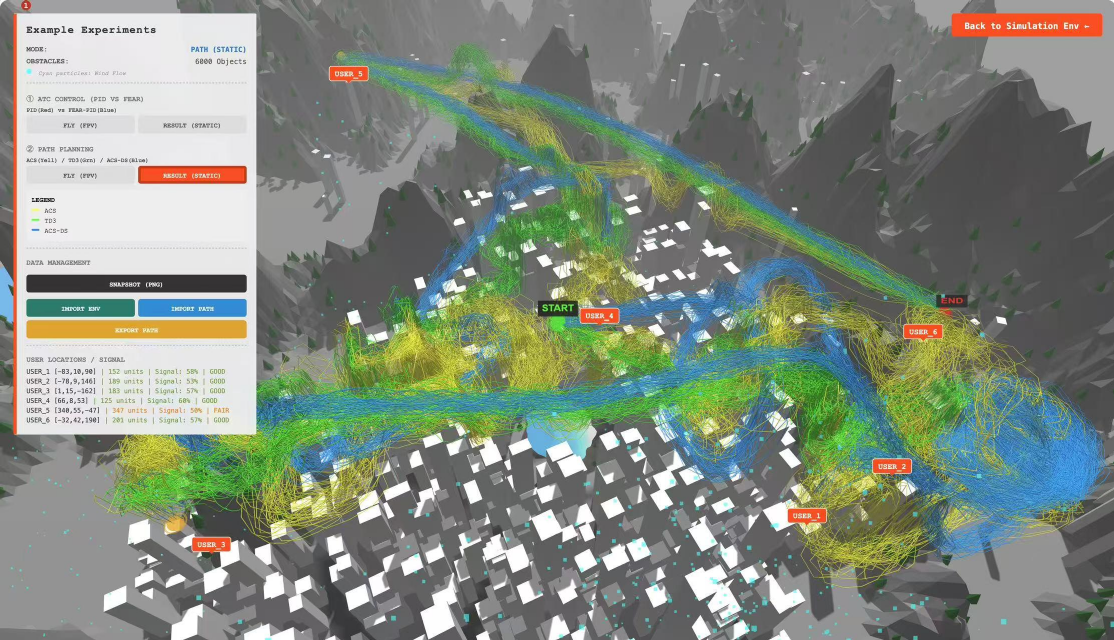}
    \caption{Trajectory comparison of ACS (yellow), TD3 (green), and ACS-DS (blue). Visualization of $50$ independent runs for evaluating trajectory stability and robustness.}
    \label{fig:trajectory_compare_acs}
\end{figure}
}
\revise{
\subsubsection{Ablation Study Results}
We summarize the ablation results in Fig.~\ref{fig:ablation_bar}, which report the average operational efficiency cost and its variance across configurations. The full model consistently achieves the lowest cost. Removing the ACS-DS trajectory planner causes the sharpest degradation, reflecting its importance in balancing energy and delay. The absence of PSO-based task assignment also leads to significant overhead due to poor task-resource coordination. 

Interestingly, even the removal of the FEAR-PID attitude controller, while seemingly less impactful than trajectory and assignment modules, causes noticeable inefficiencies, particularly in maneuvering-intensive segments such as ascent, descent, and turns. This confirms that flight stability indirectly influences task execution and communication quality. The gamma-based channel allocation strategy, though relatively lightweight, also contributes to system-wide efficiency by avoiding bandwidth contention and ensuring smoother transmission.

Overall, the results confirm that all modules are complementary. The holistic approach offers not only the best mean performance but also the most stable behavior, reinforcing the need for joint optimization in UAV-assisted fog computing.

\begin{figure}[ht]
    \centering
    \includegraphics[width=\linewidth]{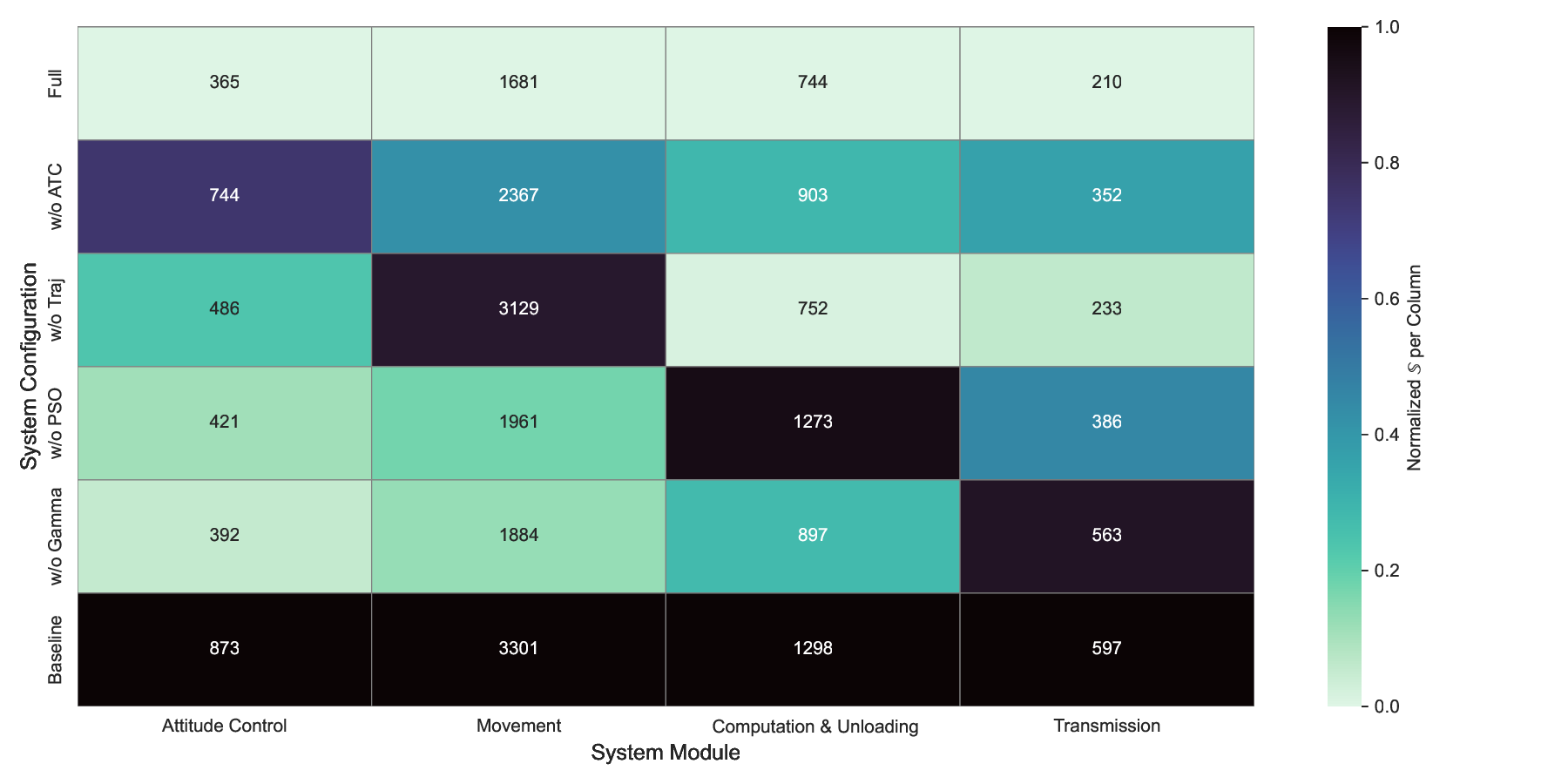}
    \caption{Average operational efficiency cost across ablation variants.}
    \label{fig:ablation_bar}
\end{figure}}

%\vspace{-0.1cm} % *****
\section{Conclusions} \label{sec: conclusion}
In this paper, we have proposed a joint optimization framework to reduce the operational efficiency cost in UAV-assisted fog computing systems. Our proposed framework involves multiple modules, including quadrotor UAV attitude control, trajectory planning in a three-dimensional spatial domain with continuously varying terrain heights, and energy-efficient assignment of computing tasks to different components in the network. We have designed appropriate mechanisms or algorithms for each module in the framework, and integrated them together to obtain a holistic solution to improve the overall efficiency in UAV-assisted fog computing. Specifically, we have proposed a novel FEAR-PID control mechanism for effective attitude control, designed the ACS-DS algorithm that overcomes the convergence issue in conventional approaches for trajectory planning in three-dimensional domains, and a modified PSO algorithm to determine the optimal task assignment. Numerical results from a wide range of experiments have shown that our proposed framework can reduce the operational efficiency cost significantly compared to existing approaches.

% \section*{Acknowledgements}
% This work is partly supported by Zhuhai Basic and Applied Basic Research Foundation Grant ZH22017003200018PWC, and partly supported by the Guangdong Provincial Key Laboratory of 
% Interdisciplinary Research and Application for Data Science, BNU-HKBU United International 
% College, Project code 2022B1212010006 and in part by Guangdong Higher Education Upgrading
% Plan (2021-2025) UIC  R0400001-22.

\appendix
%\section*{APPENDIX A} 
\section{The Dynamics of a Quadrotor UAV} \label{appendixA}
We consider two key components in the structure of a quadrotor UAV, namely the body frame and the earth frame. The body frame is commonly used for attitude control of the UAV, where the positive direction is the direction of ascent corresponding to the centers of the four motors. 

We assume that the airframe has the following characteristics and constraints,
\begin{itemize}
    \item The structure of the quadrotor UAV is symmetric;
    \item Friction between propeller and motor spindle is negligible;
    \item The stator magnetic field speed and rotor speed of the motors are infinitely close to each other, such that the slip rate is $0$;
    \item The quadrotor and propeller structures are rigid, with a uniform mass distribution and the geometric center is the mass center;
    \item The Euler angles are bounded, i.e.,  $-\pi / 2<\phi<\pi / 2, -\pi / 2<\theta<\pi / 2, -\pi<\psi<\pi$.
\end{itemize}

Based on the above assumptions, we combine the dynamics principles to establish a rigid body model for attitude control of quadrotor UAV. It should be noted that all UAV coordinate system transformations are obtained by rotation synthesis with respect to the fixed coordinate system, the rotation matrix for the conversion of the body frame to the  earth frame with the following equation,
\begin{equation}
\resizebox{\linewidth}{!}{
    $
    \begin{aligned}
    R_{b}^{e} &= \left(R_{e}^{b}\right)^{-1} = \left(R_{e}^{b}\right)^{T} = \\
    & \begin{bmatrix}
    \cos \theta \cos \psi & \cos \psi \sin \theta \sin \phi-\sin \psi \cos \phi & \cos \psi \sin \theta \cos \phi+\sin \psi \sin \phi \\
    \cos \theta \sin \psi & \sin \psi \sin \theta \sin \phi+\cos \psi \cos \phi & \sin \psi \sin \theta \cos \phi-\cos \psi \sin \phi \\
    -\sin \theta & \sin \phi \cos \theta & \cos \phi \cos \theta
    \end{bmatrix}
    \end{aligned}
    $
}
\end{equation}

Four inputs, including total thrust $U_1$, roll torque $U_2$, pitch torque $U_3$, and yaw torque $U_4$, can be controlled based on the rotation speeds of the four motors of the UAV, that is,
\begin{equation} \label{eqn: control_quantity}
\resizebox{0.9\linewidth}{!}{
    $
    \left\{\begin{array}{l}
    U_{1}(t) = C_{T}(t)\left(\omega_{1}(t)^{2}+\omega_{2}(t)^{2}+\omega_{3}(t)^{2}+\omega_{4}(t)^{2}\right) \\
    U_{2}(t) = C_{T}(t)\left(-\omega_{2}(t)^{2}+\omega_{4}(t)^{2}\right) \\
    U_{3}(t) = C_{T}(t)\left(-\omega_{1}(t)^{2}+\omega_{3}(t)^{2}\right) \\
    U_{4}(t) = C_{M}(t)\left(-\omega_{1}(t)^{2}+\omega_{2}(t)^{2}-\omega_{3}(t)^{2}+\omega_{4}(t)^{2}\right)
    \end{array}\right.
    $
}
\end{equation}

%These variables are calculated by taking the hardware parameters of the motor and propeller as inputs. Therefore, the propeller design and the method of controlling the motor output are very important.

\bibliographystyle{elsarticle-num}
\bibliography{ref}

\end{document}